\documentstyle[epsf,makeidx,bezier,subfigure,aps]{revtex}
\makeindex
\begin{document}
\baselineskip 20pt
\def\author#1{\begin{center}{\rm #1}\end{center}}
\def\address#1{\begin{center}{\it #1}\end{center}}
\def\ABSTRACT#1{
 {\centerline{\small \bf ABSTRACT}}
	\vspace{0.4cm}
	\centerline{\begin{minipage}{10cm}
		 {\small \it #1}
	\end{minipage}}
	}

\vglue 2cm
{\centerline{\bf\LARGE Analytical Study of the Julia Set}}
{\centerline{\bf\LARGE of a Coupled Generalized Logistic Map\footnote{The work is supported in part by the Grant-in-Aid for General Scientific Research from the Ministry of Education, Science, Sports and Culture, Japan (No 10640278).
}}}
\vspace{1.5 cm}
\author{Katsuhiko YOSHIDA\footnote{e-mail address: yoshida@phys.metro-u.ac.jp}\quad  and\quad Satoru SAITO\footnote{e-mail address: saito@phys.metro-u.ac.jp}}
\vspace{0.3 cm}

\address{${}^1$ School of Science, Kitasato University\\ 1-15-1 Kitasato, Sagamihara, Kanagawa, 228 Japan}

\address{${}^2$ Depertment of Physics, Tokyo Metropolitan University\\
 1-1 Minamiohsawa, Hachiohji, Tokyo 192-03 Japan }
\vspace{3 cm}

\ABSTRACT{A coupled system of two generalized logistic maps is studied. In particular influence of the coupling to the behaviour of the Julia set in two dimensional complex space is analyzed. It is proved analytically that the Julia set disappears from the complex plane uniformly as a parameter, which interpolates between chaotic phase and integrable phase, approaches to the critical value as far as the coupling strength satisfies a certain condition.}

\vspace{1 cm}
\vspace{1 cm}
\vfill\eject

\section{Introduction}

The study of nonlinear dynamical systems in the last two decades uncovered the existence of hierarchy of completely integrable models in one hand and of universal structure of nonintegrable models on the other hand. The former is characterized by concepts such as solitons, inverse methods, Yang-Baxter equations, infinite dimensional symmetries etc., whereas the terms such as chaos, fractals, strange attractors, etc. characterize the latter.

Since there exists no general method to analyze nonlinear equations the {\it completely} integrable models are quite exceptional among other nonlinear systems. Nevertheless they appear commonly in fundamental theories of physics, such as Einstein's theory of gravity, string theory of elementary particles, solvable lattice models in statistical physics and soliton models in various fields. They are easily distinguished from the behaviour of chaos solutions of nonintegrable systems. They must be also distinguished from stable solutions of nonintegrable systems.

A Julia set is a set of points of initial values on the complex plane of dependent dynamical variables whose iterational mapping never converge\footnote{This definition of a Julia set applies in one dimensional maps. Since there has been not known unique definition of Julia set in higher dimensional maps, what we call here a Julia set is somehow vague. More precise definition appropriate to our purpose will be given in \S 4.}\cite{Beardon}. A chaos will be observed experimentally if a Julia set intersects the real axis in the complex plane. Physical phenomena are expected being stable if the Julia set stays far from the real axis. The existence of a Julia set somewhere on the complex plane prevents the system to be solved analytically, besides some exceptional cases. Conversely there is no room for a Julia set to appear in the complex plane if the system is completely deterministic and predictable. From this view point the Julia set is an invariant of the dynamics, hence is a proper object which characterizes complete integrable systems among others.

Some topological aspects of Julia set have been studied for simple cases from mathematical view points\cite{Beardon}. It must possess full of information of the dynamics. Their physical significance is, however, not known. We like to emphasize here that the Julia set is very sensitive topological object against to any small difference of parameters. For instance two systems which are in chaotic orbits independently, often turn to a stable motion when they are coupled weakly. This can be understood that the Julia set is removed from the real axis as the coupling increases from zero.

In the previous papers\cite{SSSY},\cite{SSKY} we studied the behaviour of the Julia set at the point of transition from a nonintegrable phase to an integrable phase by using a simple model. It includes both the integrable and nonintegrable logistic maps as special cases of a parameter which interpolates them. This parameter enables us to study the system analytically near the critical point of transition. It was shown that the Julia set which was a Cantor set concentrated along the real axis became a countable number of discrete points as a parameter approached to the critical point of the transition\cite{SSSY}. Moreover in \cite{SSKY} we have shown that the Julia set of the generalized logistic map (GLM) converges uniformly to a certain regular orbit in the integrable limit.

The purpose of this paper is to study the behaviour of the Julia set under the influence of the coupling. To be specific we consider the generalized logistic map (GLM) as a unit of dynamical system. Our strategy of studying the coupling dependence of the Julia set is to find a model of coupling as simple as possible, so that we can see analytically the influence of the coupling. Therefore we study in this paper a system in which there are two coupled  GLM's.

When two units of GLM's are coupled, the Julia sets appear on each complex plane of the variables. Since they are correlated with each other through the coupling we are to study their behaviour in the hyper plane of two complex variables, or four real variables. 

In spite of the complexity arose from the hyper complex plane, we will show that our coupling model of two GLM's enables us to extract certain informations about the Julia set. In particular we will show that, in addition to those in the case of single map, there appear new fixed points of the map which cause a deformation of the Julia set. We will show that we can study such deformation analytically to some extents. We also prove that the Julia set converges uniformly to the orbits of integrable map, as far as the value of the coupling strength satisfies a certain constraint. 

Our paper is organized as follows. We will review some results of the case of single GLM in the next section. In \S 3 we introduce the coupled system and study its basic features. The main results are discussed in \S 4, where the coupling dependence of the Julia set is analysed in detail. In order to supply informations we will present numerical results of some projections of the Julia set, so that we can see visually the coupling dependence. The last section is devoted to discussions.

\vfill\eject
\section{Julia set of single generalized logistic map}

The Julia set itself does not depend on initial values, but is uniquely determined by dynamical equation. Since it is a highly complicated set of points in the complex plane analytical study of its behaviour has been limited to simple models\cite{Beardon}. In particular little is known about properties of Julia set in the case of coupled systems. Therefore, before studying the coupled case, it will be worthwhile to summarize properties of the Julia set of a single generalized logistic map (GLM), which we investigated in our previous work\cite{SSKY}.

\subsection{General feature of the Julia set of single GLM}

In the previous paper\cite{SSSY},\cite{SSKY} we studied a map which reduces to the logistic equation in the continuum limit and interpolates integrable and nonintegrable logistic maps:
\begin{equation}
z_{l+1}=f(z_l):=\mu{z_l(1-\gamma z_l)\over 1+\nu(1-\gamma)z_l};\quad z_l \in \mbox{\boldmath$C$},\quad l\in \mbox{\boldmath$Z$}.
\label{eqn:GLM}
\end{equation}
Here $\mu$, $\nu$ and $\gamma$  are complex parameters. When $\gamma = 1$ this is the standard logistic map. When $\gamma = 0$ it becomes a M\"obius map, hence we called it the integrable logistic map (ILM). The latter version of the logistic map was first studied by Morisita as an alternative difference analogue of the logistic equation. It has a unique solution given by
\begin{equation}
z_l={\mu^lz_0\over 1+\nu\displaystyle{1-\mu^l\over 1-\mu}z_0}
\label{eqn:solution}
\end{equation}
for an arbitrary initial value $z_0 \in \mbox{\boldmath$C$}$. The continuous time limit of $(\ref{eqn:GLM})$ can be taken by replacing 
$$
z_l={\mu-1\over\nu+\gamma\mu-\gamma\nu}u(t),\qquad \lim_{h\rightarrow 0}{{z_{l+h}-z_l}\over{h}}={\mu-1\over\nu+\gamma\mu-\gamma\nu}{d\over dt}u(t),
$$
\begin{equation}
\mu=1+ah
\label{eqn:continuous limit}
\end{equation}
in the limit of $h \rightarrow 0$, from which follows the logistic equation :
\begin{equation}
{du\over dt}=au(1-u).
\end{equation}
The merit of studying the map $(\ref{eqn:GLM})$ is that we can investigate analytically the transition of the dynamical system from nonintegrable phase to the integrable phase. 

We will use more convenient form, in the place of $(\ref{eqn:GLM})$, the standard parameterization of the rational map of rank 2 
\begin{equation}
F(z)=\phi^{-1}\circ f\circ \phi(z)={z(z+\lambda)\over 1+\lambda' z}\ e^{i\theta}\label{eqn:F(z)}
\end{equation}
\begin{equation}
\lambda=\mu e^{-i\theta},\qquad\lambda'={\nu\gamma-\nu-2\mu\gamma+\mu^2\gamma\over (\nu\gamma-\nu-\gamma)\mu} e^{i\theta},
\end{equation}
after the proper transformation :
\begin{equation}
\phi(x)={(1-\mu)x\over(\nu\gamma-\nu-\mu\gamma)x+(\nu\gamma-\nu-\gamma)\mu e^{-i\theta}}.
\end{equation}
In this form of the map ILM becomes extremely simple :
\begin{equation}
F(z)=\mu z,\qquad {\rm if}\quad \gamma=0.
\label{eqn:muz}
\end{equation}
In the new parameters the integrable limit $\gamma=0$ corresponds to $\lambda\lambda'=1$.

The most fundamental analytical objects are fixed points. The GLM has three fixed points, 0, $\infty$ and $z_p$
\begin{equation}
z_p := {{\lambda - e^{-\imath\theta}}\over{\lambda'e^{-\imath\theta} - 1}},\label{eqn:fixedpt}
\end{equation}
whose properties can be easily deduced from the values of derivative at those points. When its absolute value is less than 1, the point is attractive and when it is greater than 1, the point is repulsive.

 As it was explained in the introduction, the Julia set is an invariant of the mapping. Therefore it should provide us useful knowledge about global nature of the system.

A Julia set is a set of points of initial values whose iterational maps neither converge nor diverge. There are several equivalent methods to calculate the points. We will adopt the following definition of the Julia set\footnote{This is not the conventional definition\cite{Beardon} of the Julia set. Nevertheless we adopt this definition which is natural and more convenient to understand the transition to the integrable phase.} :

\noindent
{\it Definition :}
\begin{minipage}[t]{13cm}
{\it If $\varphi$ is a map of $z$ into $\varphi(z)$, and exists its inverse map $\varphi^{-1}$, then the Julia set $J(\varphi)$ of $\varphi$ is defined by the following set of points :}
\end{minipage}
\begin{equation}
J(\varphi) := \left\{ z | \mathop{\bigcup}^\infty_{l = 0} \varphi^{-l}( z_J ), {}^\forall z_J \in J(\varphi) \right\}.
\label{eqn:definition}
\end{equation}

If $z_J$ is a point of the Julia set, its every image belongs to the Julia set. Conversely, if points in a neighborhood of the Julia set do not belong to the Julia set, then all images spread to whole complex plane. Therefore the definition makes sense.

From our definition of Julia set described above, we can create the Julia set by considering its inverse mapping. The inverse mapping of GLM is given by
\begin{eqnarray}
z_l &=& F^{-1}(z_{l+1})\nonumber\\
&=& \frac{1}{2} ( \rho z_{l+1} - \lambda ) \pm \frac{1}{2}\sqrt{( \rho z_{l+1} - \lambda )^2 + 4z_{l+1}e^{-\imath\theta} }, 
\label{eqn:inverse GLM}
\end{eqnarray}
where we defined
\begin{equation}
\rho:=\lambda'e^{-i\theta}.
\end{equation}
Attractors of this mapping constitute the Julia set. The feature of the mapping is determined by dynamics of the system. 

It is well known that the behavior of a dynamical system, hence the nature of the Julia set, is sensitive to the values of the critical points, the zeros of the derivative of the mapping. In our problem, by solving $F'(z)=0$, they are
\begin{equation}
z_{crt} = \frac{1}{\lambda'}( -1 \pm \sqrt{ 1 - \lambda\lambda' } ).
\end{equation}
The mapping is discriminated by the value of
\begin{equation}
\epsilon:=\lambda' \lambda - 1.
\end{equation}
To be specific we consider the case of real parameters, i.e., $\lambda,\ \lambda'$ are real and set $\theta = 0$. Under these conditions, $\epsilon$ is real, and there arise three types of critical points, depending the values of $\epsilon$:

\begin{enumerate}
\item Two complex critical points ($\epsilon > 0$)\\
There is no critical point on the real axis and graph becomes monotonically increasing two curves. The inverse mapping transforms a real value to a real one. If $\lambda' > 0$, one of the fixed points $0$, which is on the real axis, is a repulsive one. Hence the whole iterational images of this point belong to the Julia set which are constrained on the real axis. In fact it becomes a Cantor set as mentioned in ref.\cite{SSSY}.

\item One critical point ($\epsilon = 0$)\\
Two critical points join into one at the singular point $z = - \lambda$, and dynamics becomes linear. In this case the Julia set has gone and system becomes integrable. We will call $\epsilon$ the integrability parameter in what follows.

\item Two real critical points ($\epsilon < 0$)\\ 
The graph has two peaks along the real axis and they can be considered as two independent monotonic systems. In this case one of the critical points in between two fixed points $0$ and $z_p$ is important. Many patterns of the Julia set appear in the complex plane and exhibit complicated structures such as Jordon curves, Cantor sets or dentrites, depending on location of the critical points.

\end{enumerate}

\subsection{Integrable limit of the Julia set of single GLM}
In our previous paper\cite{SSKY} we have shown in the case of single GLM that the Julia set converges uniformly to the set of points of the M\"obius maps in the integrable limit. We will summarize briefly the proof in this subsection.

Let us first consider the integrable case, i.e., $\epsilon=0$. If $|\lambda|>1$ the origin is a repeller. The inverse map $(\ref{eqn:inverse GLM})$ started from $z=0$ turns to be
\begin{equation}
z_l=\cases{\rho z_{l+1}\cr -\lambda\cr},
\label{eqn:ILM set}
\end{equation}
which yields an orbit consisting of the set of points 
\begin{equation}
J^{ILM}=\{-\rho^n\lambda\Bigl| n\in \mbox{\boldmath$N$}\}.
\end{equation}

We now consider the cases in which $\epsilon$ does not vanish. The inverse map $(\ref{eqn:inverse GLM})$ becomes multivalued and much complicated. Nevertheless we could prove the following fact:
\vglue 0.3cm
\centerline{{\it All points of the Julia set approach uniformly to $J^{ILM}$ in the integrable limit.}}
\vglue 0.3cm

In order to see the behavior of the Julia set at small values of $\epsilon$, it is convenient to rewrite $(\ref{eqn:inverse GLM})$ as
\begin{equation}
F^{-1}(z)=\left\{\matrix{A(z)\cr B(z)\cr}\right.\qquad A:\ z\rightarrow \rho z +E(z),\quad B:\ z\rightarrow -\lambda-E(z)
\label{eqn:F^-1}
\end{equation}
where
\begin{equation}
E(z):={1\over 2}(\rho z+\lambda)\left(\sqrt{1-{4z\epsilon 
e^{-i\theta}\over(\rho z+\lambda)^2}}\ -\ 1\right). 
\label{eqn:E(z)}
\end{equation}
$E(z)$ vanishes for small values of $\epsilon$. In fact we can show
\begin{equation}
|E(z)|\le R_\epsilon,
\label{eqn:|E(z)|<}
\end{equation}
for all values of $\epsilon$. Here $R_\epsilon$ is given by\footnote{This result, which we owe to Hisao Konuma, improves the one of our previous paper \cite{SSKY}.}
\begin{equation}
R_\epsilon={\sqrt 2+1\over|\lambda'|}\sqrt{|\epsilon|}\left(\sqrt{|\epsilon|}+\sqrt{|\epsilon+1|}\right).
\end{equation}
which tends to zero like $\sqrt{|\epsilon|}$ as $\epsilon$ approaches to zero. It is remarkable that $R_\epsilon$ does not depend on the argument of the map, but is determined only by the parameters characterizing the map.

We can prove our claim as follows:
\begin{enumerate}
\item
Using $A$ and $B$ defined in $(\ref{eqn:F^-1})$ the $n$th iteration of $F^{-1}$ yields
\begin{equation}
F^{-n}(z)=\left\{\left. A^{\nu_1}B^{\nu_2}A^{\nu_3}\cdots
 B^{\nu_n}(z)\right|\ \nu_1+\nu_2+\cdots+\nu_n=n\right\}.
\label{eqn:F^-n}
\end{equation}
\item
For any $X$ an element of the form $A^sBX,\ s=0,1,2,\cdots, n-1$ in $(\ref{eqn:F^-n})$ lies in the neighbourhood of $-\rho^s\lambda$ :
\begin{eqnarray}
|A^s(BX)+\rho^s\lambda|&=&\left|-\rho^sE(X)+\sum_{k=0}^{s-1}\rho^kE\left(A^{s-k-1}BX\right)\right|\nonumber\\
&<&\ {1-|\rho|^{s}\over 1-|\rho|}R_\epsilon,\qquad s=0,1,2,\cdots, n-1.
\label{eqn:futo}
\end{eqnarray}
Since we assumed $|\lambda|>1$, $|\rho|=|\lambda'|<1$ is satisfied as long as $\epsilon$ is small. Hence the right hand side of $(\ref{eqn:futo})$ is finite for all $s$.

The exception is the case $A^n(0)$, for which $|A^n(0)|=0$ holds since $z=0$ is a fixed point of the map $A$.
\item From 1. and 2. we conclude
\begin{equation}
\left|F^{-n}(z_0)-(-\rho^{n}\lambda)\right|<{1\over 1-|\rho|}R_\epsilon\rightarrow 0,\quad ^\forall\ n\quad \epsilon\rightarrow 0.
\end{equation}
In our present problem the Julia set is the collection of all $F^{-n}(0)$, i.e.,
\begin{equation}
J^{GLM}=\mathop{\bigcup}_{n=0}^\infty F^{-n}(z_J),\qquad z_J\in J^{GLM},
\label{eqn:J^GLM}
\end{equation}
hence the claim is justified.
\end{enumerate}

\vfill\eject
\section{Interference of two GLM's}

We now study the coupling of GLM. The most general linear coupling is given in the following form
\begin{equation}
z_{l+1}^{(j)}=\sum_{k}g_{jk}F(z_l^{(k)})
\label{eqn:cml}
\end{equation}
with arbitrary constants $\{g_{jk}\}$ and $F(z)$ being a mapping function. We should recall here the fact that the Julia set has been studied mainly from mathematical point of view, and restricted to the case of small number of variables. Since we are interested in studying influence of the coupling to the Julia set generated by the function $F$, we will restrict our discussion to the simplest case of the coupling, namely to the coupling of two units. Hence the system to be studied in the following is then
\begin{eqnarray}
z_{l+1}&=&\varphi(z_l,z'_l)=(1-g)F(z_l)+gF(z'_l),\nonumber\\
z'_{l+1}&=&\varphi(z'_l,z_l)=gF(z_l)+(1-g)F(z'_l)
\label{eqn:2 GLM}
\end{eqnarray}

\subsection{Fundamental feature of the mapping}

First let us notice some results which can be drawn from this expression by a glance. It has an invariance under the change of the coupling strength $g\ \rightarrow\ 1-g$ if the variables $z_l$ and $z'_l$ are exchanged simultaneously. Therefore it is sufficient to investigate the limitted region $g<{1\over 2}$. When $g={1\over 2}$, two equations of $(\ref{eqn:2 GLM})$ coincide, hence the map reduces to the one of single GLM.

$(\ref{eqn:2 GLM})$ is a map from two dimensional complex space on itself in general. There exists, however, a special subset of maps which close among themselves and remain essentially one dimensional. It occurs once $z_l$ and $z'_l$ coincide each other. Then, as it is clear from the expression of $(\ref{eqn:2 GLM})$, all subsequent mapping give the common values to them, and the mapping becomes equivalent to a single map. This phenomenon is not special only for $(\ref{eqn:2 GLM})$, but it happens to all linear mapping of the form $(\ref{eqn:cml})$ under the constraint $\sum_{k=1}^ng_{jk}=1,\ j=1,2,\cdots, n.$

The most basic informations are given by fixed points of the mapping. In the case of coupled maps it is not clear at all if there exists a way to find them analytically. Therefore it is remarkable that in the case of two unit coupling of GLM we can find them analytically, as we see below. This fact enables us to study dynamics of the system in detail.

Solving $(\ref{eqn:2 GLM})$ for $\mbox{\boldmath $Z$} =(z, z')$ with $(z, z')=(z_{l+1},z'_{l+1})=(z_l,z'_l)$ we find
\begin{equation}
\mbox{\boldmath $Z$}_0=(0,0),\quad \mbox{\boldmath $Z$}_p=\left(z_p,\ z_p\right),\quad \mbox{\boldmath $Z$}_\infty=(\infty, \infty), \quad \mbox{\boldmath $Z$}_\pm = (z_\pm,z_\mp),
\end{equation}
where $z_p$ is the same as $(\ref{eqn:fixedpt})$, and
\begin{equation}
z_\pm=-{1\over 2}\ {1-(1-2g)\lambda e^{i\theta}\pm\sqrt{(1-2g)^2(1-\lambda e^{i\theta})^2+4g^2\displaystyle{1-2g\over \lambda e^{-i\theta}-1+2g}(1-\lambda\lambda')}\over (1-g)\lambda'-(1-2g)e^{i\theta}},
\label{eqn:Z_pm}
\end{equation}

We notice that $\mbox{\boldmath $Z$}_0,\ \mbox{\boldmath $Z$}_p,\ \mbox{\boldmath $Z$}_\infty$ are also fixed points of the map of the single unit $F(z)$ itself\cite{SSKY}. The result is already quite interesting. When one of the units is bound to one of the fixed points of the single unit the other one is also attracted to the same point in another plane. This can be seen from the formula 
\begin{equation}
z'={1-g\over g}z-{1-2g\over g}F(z)
\label{eqn:(z',z)}
\end{equation}
which follows to $(\ref{eqn:2 GLM})$. Therefore it is forbidden that, for instance, one falls to the attractor at $z=0$ and the other falls to $z=\infty$. On the other hand $\mbox{\boldmath $Z$}_\pm$ are new and if one of GLM's is attracted to $z=z_+$, the other one must falls to $z_-$, and vice versa.

In the neighbourhood of a fixed point the behaviour of the mapping is characterized by the nature of the Jacobi matrix and its eigenvalues. The general expressions of the Jacobi matrix and the eigenvalues associated with our model are given in Appendix. There we also present illustration of the regions of parameters which correspond to repulsive, attractive and saddle points.

We will summarize, in the following,  properties of the fixed points.

Eigenvalues for $\mbox{\boldmath $Z$}_0$ and $\mbox{\boldmath $Z$}_p$ are
\begin{equation}
\Lambda_0= \lambda e^{i\theta}(1-g\pm g),\qquad 
\Lambda_p = {2-\lambda e^{i\theta}-\lambda' e^{-i\theta} \over 1-\lambda\lambda'}e^{i\theta}(1-g\pm g).
\label{eqn:Lambda_0,p}
\end{equation}
They agree with those of uncoupled map when $g=0$.

In the neighbourhood of infinity the eigenvalues are 
\begin{equation}
\Lambda_\infty={\lambda'e^{-i\theta}\over 2}\left[(1-g)(a^2+b^2)\pm\sqrt{(1-g)^2(a^2-b^2)^2+4g^2a^2b^2}\right],
\label{eqn:LAMBDA_infty}
\end{equation}
where
$$
a={1\over 1-g+rg},\quad b={1\over 1-g+{1\over r}g},\qquad r:=\lim_{|z|,|z'|\rightarrow\infty}{z\over z'}.
$$
Notice that the parameter $r$ depends on the direction along which the limits $|z|\rightarrow\infty$ and $|z'|\rightarrow\infty$ are taken. Therefore the behaviour of the map at infinity is not determined uniquely but varies sensitively to the direction when there exists the coupling $g$. When we choose $r=\pm 1$, the result is simplified and get 
\begin{eqnarray}
\Lambda_\infty&=&\lambda' e^{-i\theta}(1-g\pm g),\quad {\rm if}\ r=1, \nonumber\\
               &=&\lambda' e^{-i\theta}{1-g\pm g\over (1-2g)^2},\qquad {\rm if}\ r=-1.
\end{eqnarray}

$\Lambda_\pm$ associated with $\mbox{\boldmath $Z$}_\pm$ do not have compact expressions but are given by
\begin{eqnarray}
\Lambda_\pm&=&{e^{i\theta}\over\lambda'}(1-g)(1-K)\nonumber\\
&\pm &{e^{i\theta}\over\lambda'}\sqrt{(1-g)^2\left[K^2-\left({\lambda' e^{-i\theta}-1+2g\over 1-2g}\right)^2\right]-g^2\left[1+\left({\lambda' e^{-i\theta}-1+2g\over 1-2g}\right)^2-2K\right]}.\nonumber\\
\label{eqn:Lambda pm}
\end{eqnarray}
\begin{equation}
K={1\over 2}\left({1-\lambda\lambda'\over (1+\lambda' z_+)^2}+{1-\lambda\lambda'\over (1+\lambda' z_-)^2}\right)
\end{equation}
This result is new since there is no correspondence in the uncoupled map model.

Given above informations we are ready to draw some pictures about the behaviour of the system under a change of the coupling. We observe the following:
\begin{enumerate}
\item
Near the fixed points $\mbox{\boldmath $Z$}_0,\ \mbox{\boldmath $Z$}_p,\ \mbox{\boldmath $Z$}_\infty$ one of two units behaves exactly the same as one of uncoupled GLM. The other unit is attracted (or repelled) at the same fixed point. This explaines the coherent behaviour often observed in a coupled system. The multiplier of the first unit remains the same but the other one changes its scale by the factor $1-2g$. Therefore the second unit increases its stability if $0 < g < 1$ and instability if $-1 < g < 0$.
\item
The fixed points at $\mbox{\boldmath $Z$}_\pm$ have no correspondence to the uncoupled GLM. When the coupling is weak, $(\ref{eqn:Z_pm})$ behave as
\begin{eqnarray}
z_+&=&z_p+g{1-\lambda\lambda'\over(1-\lambda' e^{-i\theta})^2}e^{-i\theta}+o(g^2)\nonumber\\
z_-&=&\qquad g{1\over 1-\lambda' e^{-i\theta}}e^{-i\theta}+o(g^2),
\label{eqn:z_pm +g}
\end{eqnarray}
hence $z_+$ and $z_-$ approach to $z_p$ and $0$ respectively. From this we observe that the two units behave differently if one falls to $z_\pm$. 
\item
The stability at $z_\pm$ is controlled by $|\Lambda_+|$ and $|\Lambda_-|$ respectively. In particular both of them are stable as far as $|\Lambda_\pm|$ is less than 1. In the weak coupling limit the multipliers $\Lambda_\pm$ reduce to 
\begin{equation}
\Lambda_\pm\ \rightarrow\ {1\over 2}{1\over1-\lambda\lambda'}\left[ 2-\lambda' e^{-i\theta}-\lambda^2\lambda' e^{i\theta} \pm \left(1-\lambda e^{i\theta}\right)\left(2-\lambda' e^{-i\theta}-\lambda\lambda'\right)\right],
\end{equation}
which coincide with $\Lambda_p$ and $\Lambda_0$ evaluated at $g=0$. 
\end{enumerate}
\subsection{Linear coupling of ILM}

Before studying the coupled GLM, we like to see the effect of the coupling in the case of integrable logistic map (ILM). The mapping is given by
\begin{equation}
\left(\matrix{z_{l+1}\cr z'_{l+1}\cr}\right)=\lambda e^{i\theta}\sigma^{-1}\left(\matrix{z_l\cr z'_l\cr}\right),\qquad \sigma^{-1}:=\left(\matrix{1-g&g\cr g& 1-g\cr}\right).
\label{eqn:linear map}
\end{equation} 
Since the map is linear in $z_l$, it is invertible as long as the matrix $\sigma^{-1}$ is regular. Therefore no Julia set arises in this coupling. The matrix $\sigma^{-1}$ is singular when $g={1\over 2}$. In this special case the map reduces to the single ILM.

An orbit of the map $(\ref{eqn:linear map})$ can be seen easily if we write $\sigma^{-1}$ as
\begin{equation}
\sigma^{-1}=1-2gP,\qquad P:={1\over 2}\left(\matrix{1&-1\cr -1&1\cr}\right)
\label{eqn:proj}
\end{equation}
and notice that $P$ is a projection matrix satisfying $P^2=P$. Every pair of points $\left(\matrix{z\cr z'\cr}\right)$ is mapped to the plane specified by $z+z'=0$. Then the $n$th map started from $\left(\matrix{z_0\cr z'_0\cr}\right)$ yields
\begin{eqnarray}
\left(\matrix{z_n\cr z'_n\cr}\right)&=&\lambda^ne^{in\theta}\Bigl(1+\left[(1-2g)^n-1\right]P\Bigr)\left(\matrix{z_0\cr z'_0\cr}\right)\nonumber\\
&=&\lambda^ne^{in\theta}\left(\matrix{z_0\cr z'_0\cr}\right)
+\lambda^ne^{in\theta}\left[(1-2g)^n-1\right]\left(\matrix{z_0-z'_0\cr z'_0-z_0\cr}\right).
\end{eqnarray}

The behaviour of this map is governed by the eigenvalues of the transformation. Corresponding to the eigenvalues $\lambda_\pm=\lambda e^{i\theta}(1-g\pm g)$ the eigenvectors of this map are $\left(\matrix{1\cr \pm 1\cr}\right)$. They converge either to the origin or to infinity, depending whether $|\lambda_\pm|<1$ or $>1$. These two points are nothing but the fixed points of the map of $(\ref{eqn:linear map})$.

\vfill\eject
\section{Julia set of coupled GLM}

We are interested in studying the change of the Julia set when the coupling is introduced. It should supply useful information about the effect of the coupling of GLM's. The Julia set in the case of coupled GLM, however, must be rather complicated, since two complex variables, hence four real variables, are to be considered simultaneously. Moreover a proper definition of the Julia set itself is not clear in the case of hyper complex plane. Therefore we must introduce some definition which is appropriate in our discussion. For this purpose, in this paper, we propose to call a Julia set an object which is defined simply by generalizing $z$ in $(\ref{eqn:definition})$ to a point in the hyper plane $( z, z' )$. That is,

{\it
the Julia set $J(\varphi)$ of the map $(\ref{eqn:2 GLM})$ is a closure of the inverse images }
\begin{equation}
\left\{\left(\left.\matrix{\varphi^{-n}(z_J,z'_J)\cr \varphi^{-n}(z'_J,z_J)\cr}\right)\ \right|\ n\in \mbox{\boldmath{$N$}}\right\}.
\end{equation}
{\it of an arbitrary repulsive fixed point $\left(\matrix{z_J\cr z'_J\cr}\right)$.}

The set of these points has apparently the following properties,
\begin{enumerate}
\item invariance under the map,
\item starting from any point of the set the forward map never converges to an attractor.
\end{enumerate}

The inverse map becomes much more complicated compared with the one of single GLM. In the case of the linear coupling in the form of $(\ref{eqn:2 GLM})$, however, we can study behaviour of the map analytically to some extent. We present in this section how the Julia set of the coupled system will disappear when parameters approach to the integrable limit. We also investigate the influence of the coupling to the Julia set. Our arguments will be supported by displaying deformation of the Julia set using computer simulation. 

\subsection{Inverse map of the coupled GLM}

\subsubsection{General feature of the inverse map of the coupled GLM}

Generally speaking it is a difficult problem to find inverse map of a coupled nonlinear system if it dedends on more than two variables. In the case of the coupled GLM $(\ref{eqn:GLM})$, however, we can solve this problem rather easily and obtain the following compact expression:
\begin{eqnarray}
z_l&=&{1\over 2}\left(\rho x_{l+1}-\lambda\right)\pm {1\over 2}\sqrt{\left(\rho x_{l+1}-\lambda\right)^2+4x_{l+1}e^{-i\theta}},\nonumber\\
z'_l&=&{1\over 2}\left(\rho y_{l+1}-\lambda\right)\pm {1\over 2}\sqrt{\left(\rho y_{l+1}-\lambda\right)^2+4y_{l+1}e^{-i\theta}},
\label{eqn:inverse map}
\end{eqnarray}
where
\begin{equation}
\left(\matrix{x_l\cr y_l}\right)=\sigma\mbox{\boldmath$Z$}_l,\qquad \mbox{\boldmath$Z$}_l:=\left(\matrix{z_l\cr z'_l}\right),\qquad \sigma:={1\over 1-2g}\left(\matrix{1-g&-g\cr -g&1-g}\right).
\label{eqn:inv coupled map}
\end{equation}
From this expression we see that a pair of points $(z_{l+1},z'_{l+1})$ are inversely mapped to four pairs of points. We must take into account all of such pairs at every step of the map.

There are four types of mapping which we represent as
\begin{eqnarray}
&&M_{(1)}:=\left(\matrix{A&0\cr 0&A\cr}\right)\sigma,\qquad 
M_{(2)}:=\left(\matrix{A&0\cr 0&B\cr}\right)\sigma,\nonumber\\ 
&&M_{(3)}:=\left(\matrix{B&0\cr 0&A\cr}\right)\sigma,\qquad 
M_{(4)}:=\left(\matrix{B&0\cr 0&B\cr}\right)\sigma.
\label{eqn:M_(j)'s}
\end{eqnarray}
Here $A$ and $B$ are the operations which map $z$ to $A(z)$ and $B(z)$, respectively, according to $(\ref{eqn:F^-1})$. A single operation of the inverse map to a point, say $\mbox{\boldmath$Z$}=\left(\matrix{z\cr z'\cr}\right)$, yields the four pairs of points
\begin{equation}
M_{(j)}\mbox{\boldmath$Z$},\qquad j=1,2,3,4.
\end{equation}
If we apply this map twice we get $16$ points,
\begin{equation}
M_{(j_1,j_2)}\mbox{\boldmath$Z$}:=M_{(j_1)}M_{(j_2)}\mbox{\boldmath$Z$},\qquad j_1,j_2=1,2,3,4.
\end{equation}
Similarly if we apply the same map $n$ times, we will get $4^n$ points
\begin{equation}
M_{(j_1,j_2,\cdots,j_n)}\mbox{\boldmath$Z$}:=M_{(j_1)}M_{(j_2)}\cdots M_{(j_n)}\mbox{\boldmath$Z$},\qquad j_1,j_2,\cdots,j_n=1,2,3,4.
\label{eqn:n-th map}
\end{equation}
If we start the map from a repeller ${\mbox{\boldmath{$Z$}}}_J$, the union of all these points forms the Julia set, which we represent as 
\begin{equation}
J^{(g, \epsilon)}=\mathop{\bigcup}^\infty_{n=0}\left(\mathop{\bigcup}_{j_1,j_2,\cdots,j_n}M_{(j_1,j_2,\cdots,j_n)}\mbox{\boldmath$Z$}_J\right).
\end{equation}
Here $\mathop{\bigcup}_{j_1,j_2,\cdots,j_n}$ means the union of all possible sets of $(j_1,j_2,\cdots,j_n).$
\subsubsection{`Julia set' of the coupled GLM at integrable limit}

It is instructive to study first the case of integrable coupled system, i.e., $\epsilon =0$. For this purpose we write $M_{(j)}$'s in $(\ref{eqn:M_(j)'s})$, as
\begin{equation}
M_{(j)} =M_{(j)}^{(0)}+E_{(j)},\qquad j=1,2,3,4,
\label{eqn:M=M^0+E_j}
\end{equation}
where
\begin{equation}
M_{(1)}^{(0)}:=\rho\sigma,\quad
M_{(2)}^{(0)}:=\left(\matrix{\rho&0\cr 0&-\hat\lambda\cr}\right)\sigma,\quad
M_{(3)}^{(0)}:=\left(\matrix{-\hat\lambda&0\cr 0&\rho\cr}\right)\sigma,\quad
M_{(4)}^{(0)}:=-\hat\lambda\sigma,
\label{eqn:M_j^(0)}
\end{equation}
with $\hat\lambda$ being the map which brings an arbitrary point to the fixed point $\lambda$, and
$$
E_{(1)}:=\left(\matrix{E&0\cr 0&E\cr}\right)\sigma,\qquad 
E_{(2)}:=\left(\matrix{E&0\cr 0&-E\cr}\right)\sigma,
$$
\begin{equation}
E_{(3)}:=\left(\matrix{-E&0\cr 0&E\cr}\right)\sigma,\qquad
E_{(4)}:=-\left(\matrix{E&0\cr 0&E\cr}\right)\sigma.
\end{equation}

In the $\epsilon\rightarrow 0$ limit, $E_{(j)}$'s vanish and the `Julia set' is simplified\footnote{This is not a Julia set in the sense of conventional definition\cite{Beardon}. It makes sense only by the definition we adopted in $(\ref{eqn:definition})$. Hence we call this `Julia set' to distinguish from others.} 
\begin{equation}
J^{(g, 0)}=\mathop{\bigcup}^\infty_{n=0}\left(\mathop{\bigcup}_{j_1,j_2,\cdots,j_n}M^{(0)}_{(j_1,j_2,\cdots,j_n)}\mbox{\boldmath$Z$}_J\right).
\end{equation}
Namely in $(\ref{eqn:M_(j)'s})$ we replace $A$ by $\rho$ and $B$ by $-\hat\lambda$. Notice that this result is different from what we expect naively as we take the inverse map of $(\ref{eqn:linear map})$. The latter corresponds to the $\rho\sigma$ term in $(\ref{eqn:M_j^(0)})$. The other contributions could be found only as we approach to the integrable limit from non-integrable region $\epsilon\ne 0$. This phenomenon was already observed when we studied the single GLM\cite{SSKY}. The situation is much more complicated in the present problem because of the existence of the coupling.

It is not difficult to understand the reason of this happening. In the integrable limit the fixed point $\mbox{\boldmath$Z$}_p=(z_p, z_p)$ moves to $(-\lambda, -\lambda)$ and becomes super repulsive (hence super attractive of the inverse map) in the sense that its multiplier turns to infinity. The operator $\hat\lambda$ in $M_{(2)}, M_{(3)}, M_{(4)}$ of $(\ref{eqn:M_(j)'s})$ brings every point to this particular pair of points. 

\subsubsection{Transition of the Julia set to `Julia set' in the integrable limit}

We now proceed to show that as $\epsilon$ approaches to zero a similar phenomena takes place as in the single GLM and the Julia set $J^{(g,\epsilon)}$ converges uniformly to $J^{(g,0)}$, if the parameters satisfy certain conditions. The proof goes as follows.

Let us assume $|\Lambda_0|>1$, where $\Lambda_0$ is the one of $(\ref{eqn:Lambda_0,p})$. This is satisfied if
\begin{equation}
|\lambda|>\cases{\quad 1,\quad\qquad {\rm if}\ g<0\cr
{1\over 1-2g},\qquad {\rm if}\ 0<g<{1\over 2}\cr}.
\label{eqn:Lambda_0>1}
\end{equation}
Then $\left(\matrix{z_J\cr z'_J\cr}\right)=\mbox{\boldmath$Z$}_0$ is a repeller and on the Julia set. Starting from this point and applying $n$ times the inverse map we obtain $4^n$ points of the Julia set $M_{(j_1,j_2,\cdots,j_n)}\mbox{\boldmath$Z$}_0$ to every set of $(j_1,j_2,\cdots, j_n)$. 

We are interested in the difference of $M_{(j_1,j_2,\cdots,j_n)}\mbox{\boldmath$Z$}_0$ from $M^{(0)}_{(j_1,j_2,\cdots,j_n)}\mbox{\boldmath$Z$}_0$. The latter is obtained from the former by simply erasing $E_j$'s in $(\ref{eqn:M=M^0+E_j})$ everywhere. The difference, therefore, has to contain $E$ at least once. We recall that once this function appears in the sequence of the map, all points are mapped inside of the disc of radius $R_\epsilon$. This also implies that only the last $E$ in the map is sufficient to be considered, because the other $E$'s in the past are included as informations within the disk which we do not care anyway. 

We further notice that $\hat\lambda$ terms in $M^{(0)}_{(j)}$'s reset all informations in the past and put to the fixed value $\lambda$. Hence if such term appears in the series of maps, it is exactly cancelled in the difference and we are left with only contributions of $\rho$ terms. Accordingly the difference takes the form
\begin{equation}
M_{(j_1,j_2,\cdots,j_n)}\mbox{\boldmath$Z$}_0-M^{(0)}_{(j_1,j_2,\cdots,j_n)}\mbox{\boldmath$Z$}_0
=\sum_{k=1}^n
 C_{j_n}\sigma C_{j_{n-1}}\sigma \cdots C_{j_{k+1}}\sigma E_{(j_k)}\mbox{\boldmath{$X$}},
\end{equation}
where we used the notations
\begin{equation}
C_1:=\rho\left(\matrix{1&0\cr 0&1\cr}\right),\quad
C_2:=\rho\left(\matrix{1&0\cr 0&0\cr}\right),\quad
C_3:=\rho\left(\matrix{0&0\cr 0&1\cr}\right),\quad
C_4=0.
\end{equation}

Let $\left(\matrix{z_{n,+}^{(\epsilon)}\cr {z}_{n,-}^{(\epsilon)}\cr}\right)$ and $\left(\matrix{z_{n,+}^{(0)}\cr {z}_{n,-}^{(0)}\cr}\right)$ be one of such pair of points associated with a particular combination of $(j_1,j_2,\cdots,j_n)$. We calculate distances of upper and lower components in the complex planes separately. They can be written as
\begin{equation}
\left(\matrix{|z_{n,+}^{(\epsilon)}-z_{n,+}^{(0)}|\cr
      |z_{n,-}^{(\epsilon)}-z_{n,-}^{(0)}|\cr}\right)
=\left|\sum_{k=1}^n\ C_{j_n}\sigma C_{j_{n-1}}\sigma\cdots C_{j_{k+1}}\sigma E_{(j_k)}\mbox{\boldmath{$X$}}
\right|
\end{equation}
Each term in the summation over $k$ contributes the factor $\rho^{n-k}$. Let us introduce the notation $\sigma_m$ to denote the maximum of the absolute value of the eigenvalues of $\sigma$, that is
\begin{equation}
\sigma_m:={\rm Max}\left\{1,\ {1\over |1-2g|}\right\}.
\end{equation}
Then we can prove
\begin{eqnarray}
\left|z_{n,\pm}^{(\epsilon)}-z_{n,\pm}^{(0)}\right|
&\le&
\sum_{k=1}^n|\rho\sigma_m|^{n-k}|E|\nonumber\\
&\le&
{1-|\rho\sigma_m|^n\over 1-|\rho\sigma_m|}R_\epsilon,
\label{eqn:limit}
\end{eqnarray}
if we replace $C_j,\ j=1,2,3$ by $\rho\left(\matrix{1&0\cr 0&1\cr}\right)$. 

This result shows that the coefficient of $R_\epsilon$ is finite for all $n$ if $|\rho\sigma_m|$ is less than 1.
 When $g<0$ or there is no coupling, $\sigma_m=1$ and the condition reduces to the case already studied for the uncoupled map. On the other hand, if $g$ is positive, $\sigma_m$ is larger than 1. In this case the condition becomes $|\lambda'|<|1-2g|$. Therefore we can say that the Julia set $J^{(g,\epsilon)}$ converges to $J^{(g,0)}$ uniformly if the parameters satisfy, together with $(\ref{eqn:Lambda_0>1})$, the following conditions:
\begin{equation}
|\lambda'|<\cases{\quad 1,\qquad\qquad {\rm if}\ g\le 1\cr |1-2g|,\qquad {\rm if}\ 0<g<{1\over 2}\cr}.
\label{eqn:rhosigma<1}
\end{equation}

A few remarks are in order.
\begin{enumerate}
\item This condition $(\ref{eqn:rhosigma<1})$ is not necessary but only sufficient for the uniform convergence.
\item In contrast to the uncoupled case the condition $|\rho\sigma_m|<1$ may not be automatically satisfied when $\left|\Lambda_\infty\right|<1$. As we see from the expression for $\Lambda_\infty$ in $(\ref{eqn:LAMBDA_infty})$, it is not easy to signify the condition for the infinity to be an attractor of the map.
\end{enumerate}

In order to see visually how the dependence of the Julia set on the integrability parameter $\epsilon$ looks like, we will present computer simulation of the inverse mapping for some values of $\epsilon$ and fixing other parameters. In the case $\left|\Lambda_0\right|>1$, $z=0$ is a repulsive fixed point. Hence the Julia set can be generated starting from there. Note that in order to visualize it in a finite region of a graph we must perform the calculation under the condition $|\Lambda_\infty|<1$, so that we do not fail to see any point from our view by adjusting the scale. 

Since the Julia set of our system is in two dimensional complex space $(z,z')$, we are not able to present its total figure in one graph. The best we can do is to present its certain projections. Here we will show projections of the Julia set to two dimensional real space $(Re(z),\ Re(z'))$. The simulations are executed at the coupling parameter $g = 0.3$ and $\lambda = 4.0$. The following pictures represent the projections corresponding to various values of the integrability parameter $\epsilon$. Here the scale of windows is $6 \times 6$ and the center of the windows is at $(Re(z),\ Re(z'))=( -2, -2 )$.

\begin{minipage}{17cm}
\vglue 0.5cm\noindent{\bf Fig.1 Julia set projected on $Re(z)-Re(z')$-plane}\hfill\break
\\
%
\begin{minipage}{8cm}
\epsfxsize=8cm \epsfysize=8cm \epsfbox{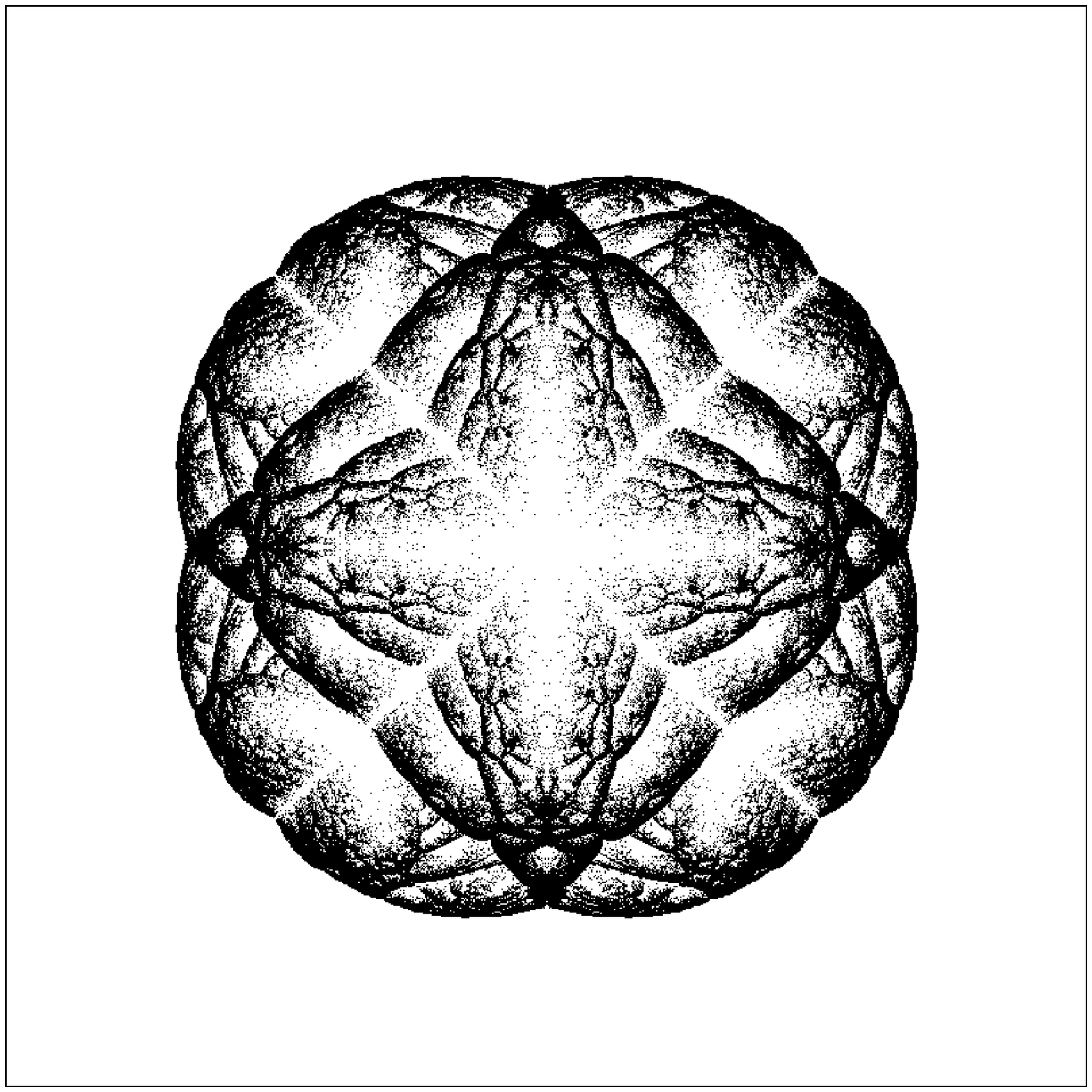}
$\epsilon=-1$
\end{minipage}
%
\begin{minipage}{8cm}
\epsfxsize=8cm \epsfbox{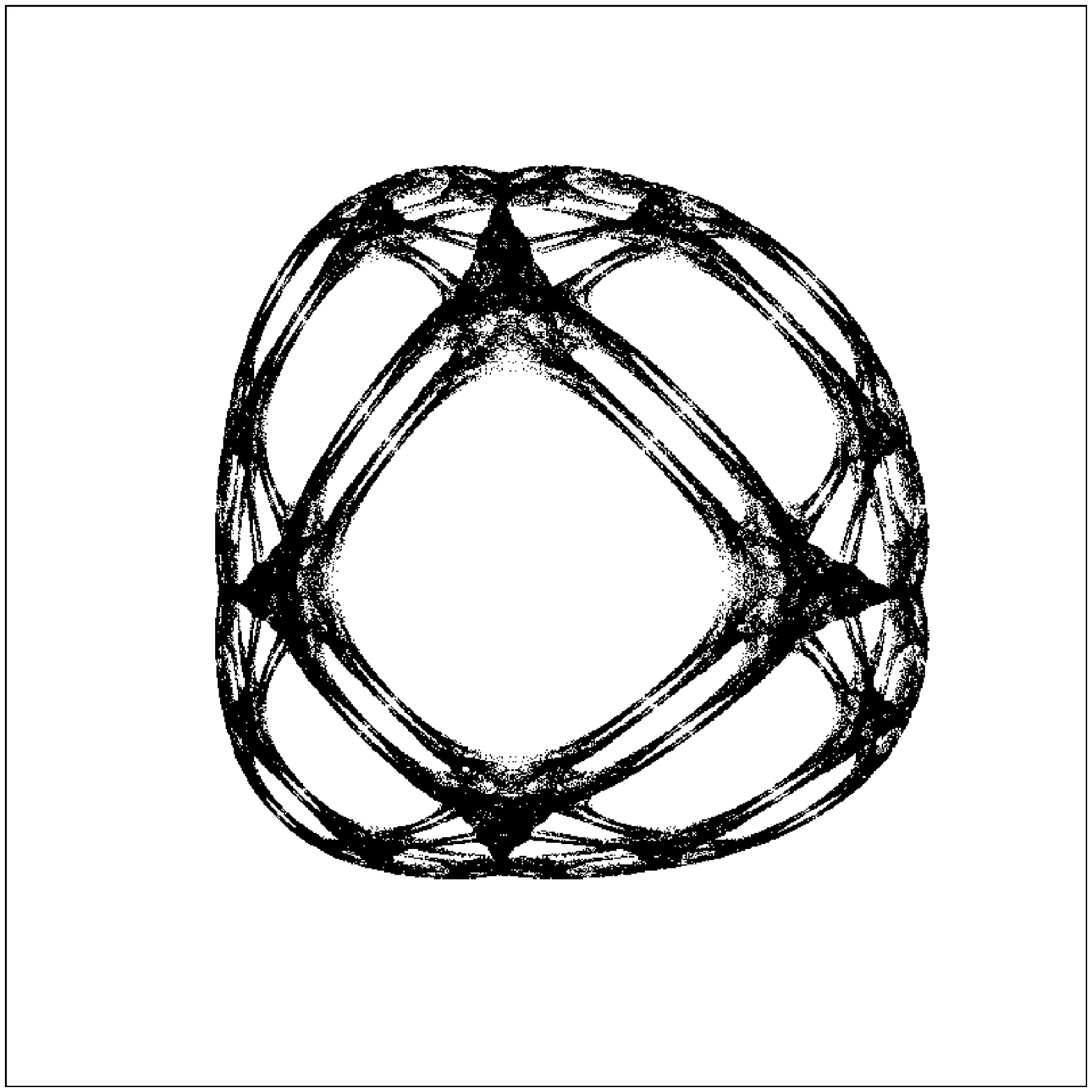}
$\epsilon=-0.6$
\end{minipage}
\\
\vglue .5cm\noindent
%
\begin{minipage}{8cm}
\epsfxsize=8cm \epsfbox{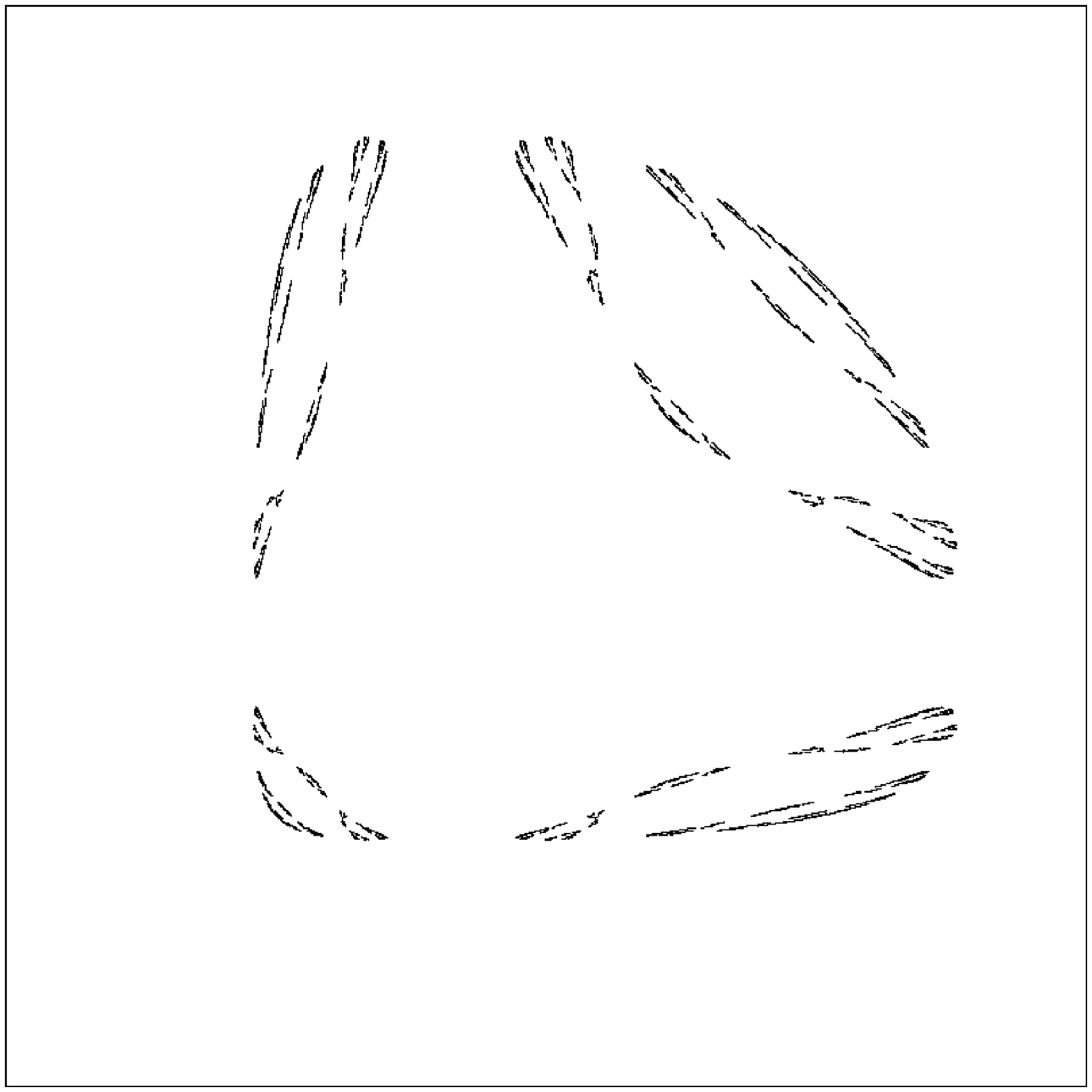}
$\epsilon=-0.2$
\end{minipage}
%
\begin{minipage}{8cm}
\epsfxsize=8cm \epsfbox{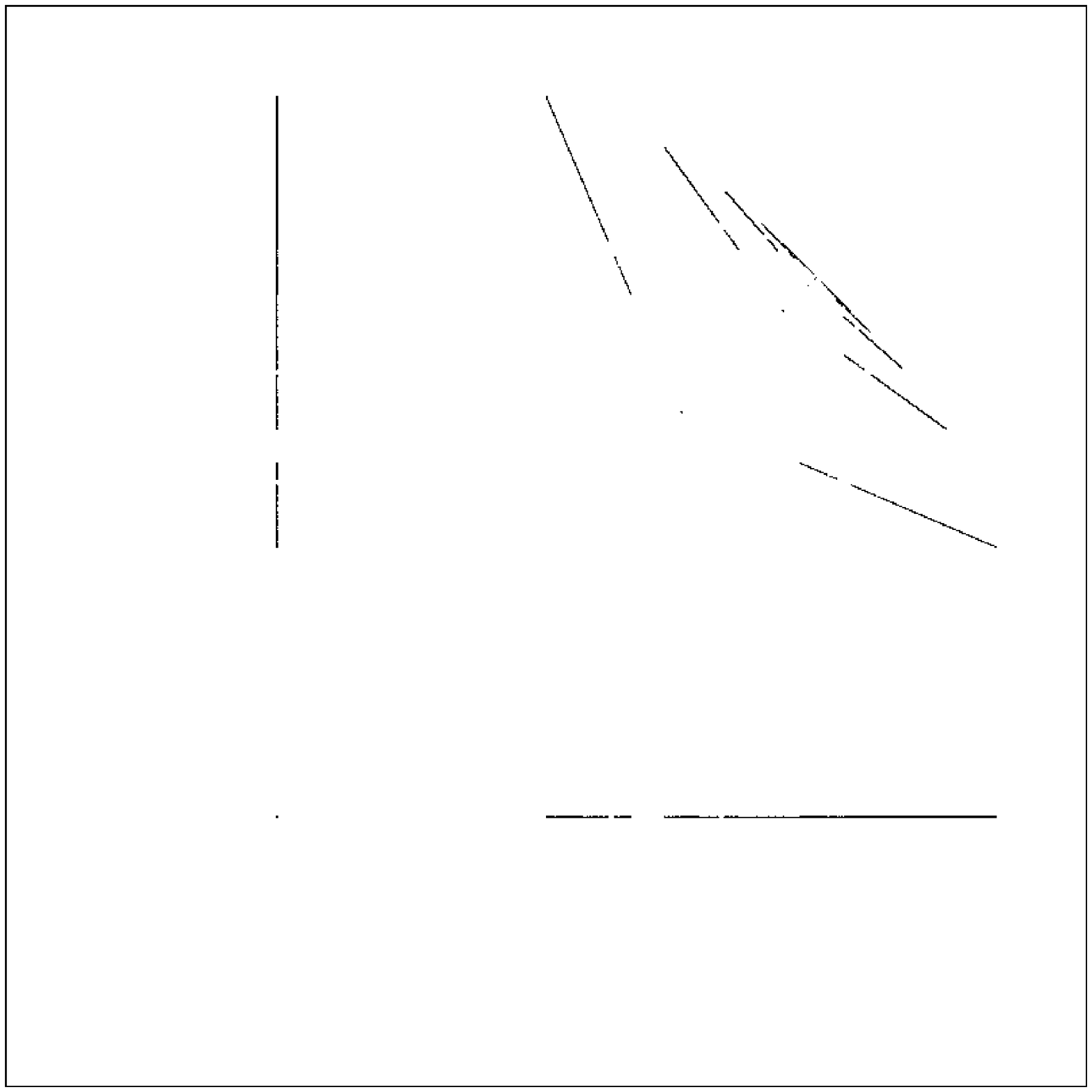}
$\epsilon=0$
\end{minipage}
\end{minipage}

\subsection{Dependence on the coupling strength}

Now we investigate the dependence of the Julia set on the coupling strength. As we see in $(\ref{eqn:inverse map})$ the inverse map depends on the coupling strength $g$ in rather complicated fashion. If we use the variables $\left(\matrix{x\cr y}\right)=\sigma\mbox{\boldmath{$Z$}}$ instead of $\mbox{\boldmath{$Z$}}=\left(\matrix{z\cr z'}\right)$, however, the dynamics becomes much simpler. In fact in the $(x,y)$ frame the coupling dependence appears in the map only through $\sigma$ given explicitly in $(\ref{eqn:M_(j)'s})$. We can employ this frame to analyze the coupling dependence. The effect of the coupling to the $(z, z')$ variables can be obtained if we transform the results of the $(x,y)$ variables to those of $(z,z')$ by $\sigma^{-1}$.

In order to separate the coupling dependence let us write $\sigma$ as
\begin{equation}
\sigma=1+\alpha P,\qquad \alpha:={2g\over 1-2g},
\end{equation}
$P$ is the projection operator defined in $(\ref{eqn:proj})$. The projection operator $P$ has a peculiar nature. Namely it maps any pair of points in the space of two variables $\left(\matrix{z\cr z'\cr}\right)$ to a symmetric pair of points different only by their signs. Hence $\alpha P$ maps $\left(\matrix{z\cr z'\cr}\right)$ to $\left(\matrix{\alpha(z-z')\cr \alpha(z'-z)\cr}\right)$. 

\subsubsection{Maps of the form $M_{(j)}${\it X}}

Starting from the repulsive fixed point $\mbox{\boldmath{$Z$}}_0$, we apply $n$ times the map $M_{(j)}$ and see the projection of the image on the $z$ plane. There should be $4^n$ points, as given by $(\ref{eqn:n-th map})$, if there is no degeneracy. Before going into detail let us first see how the single operations $M_{(j)},\ j=1,2,3,4$ map an arbitrary coordinate $\mbox{\boldmath{$X$}}=\left(\matrix{x\cr x'\cr}\right)$ on the two dimensional complex space $\left(\matrix{z\cr z'\cr}\right)$. From the definition $(\ref{eqn:M_(j)'s})$ we find\begin{eqnarray}
M_{(1)}\mbox{\boldmath{$X$}}&=&
\left(\matrix{A&0\cr 0&A\cr}\right)\sigma\left(\matrix{x\cr x'\cr}\right)=\left(\matrix{\rho(x+\alpha\xi)+E(x+\alpha\xi)\cr \rho(x'-\alpha\xi)+E(x'-\alpha\xi)\cr}\right),
\label{eqn:M_(1)}
\\
M_{(2)}\mbox{\boldmath{$X$}}&=&
\left(\matrix{A&0\cr 0&B\cr}\right)\sigma\left(\matrix{x\cr x'\cr}\right)
=\left(\matrix{\rho(x+\alpha\xi)+E(x+\alpha\xi)\cr -\lambda-E(x'-\alpha\xi)\cr}\right),\\
M_{(3)}\mbox{\boldmath{$X$}}&=&
\left(\matrix{B&0\cr 0&A\cr}\right)\sigma\left(\matrix{x\cr x'\cr}\right)
=\left(\matrix{-\lambda-E(x+\alpha\xi)\cr \rho(x'-\alpha\xi)+E(x'-\alpha\xi)\cr}\right),\\
M_{(4)}\mbox{\boldmath{$X$}}&=&
\left(\matrix{B&0\cr 0&B\cr}\right)\sigma\left(\matrix{x\cr x'\cr}\right)
=\left(\matrix{-\lambda-E(x+\alpha\xi)\cr -\lambda-E(x'-\alpha\xi)\cr}\right)
\label{eqn:M_(4)}
\end{eqnarray}
where we defined $\xi:=x-x'$. From this result we learn that the operation of $B$ maps all coordinates into the neighbourhood of the fixed point $-\lambda$ with radius $R_\epsilon$. Since $\alpha$ is only in the function $E$, the coupling dependence will not be seen unless we watch deep inside of the disk. In this sense the operation of $B$ resets the large part of informations in the past. The operation of $A$, on the other hand, depends linearly on $\alpha$. Multiple operation of $A$ will deform the Julia set depending on the strength of the coupling $g$. 

\subsubsection{Maps of the form $M_{(1)}^n({{\bf Z}}_0)$}

In order to see how the Julia set is changed from $J^{(0,\epsilon)}$, the case of uncoupled map, we study behaviour of $M_{(1)}^n{\mbox{\boldmath{$Z$}}}_0$ in detail.

If ${\mbox{\boldmath{$Z$}}}_k=\left(\matrix{z_k\cr z'_k\cr}\right)$ is the image of ${\mbox{\boldmath{$Z$}}}_0$ after $k$ times of the map, $(\ref{eqn:M_(1)})$ tells us that
\begin{equation}
\left(\matrix{z_k\cr z'_k\cr}\right)=
\left(\matrix{\rho(z_{k-1}+\alpha\xi_{k-1})+E_{k-1}\cr \rho(z'_{k-1}-\alpha\xi_{k-1})+E'_{k-1}\cr}\right)
\end{equation}
where $\xi_j:=z_j-z'_j$ and $E_j:=E(z_j+\alpha\xi_j),\ E'_j:=E(z'_j-\alpha\xi_j)$. Solving the recurrence formula
\begin{equation}
\xi_k=z_k-z'_k=\rho(1+2\alpha)\xi_{k-1}+E_{k-1}-E'_{k-1}\end{equation}
we find
\begin{equation}
\xi_n=\rho^n(1+2\alpha)^n\xi_0+\sum_{k=0}^{n-1}\rho^k(1+2\alpha)^k\left(E_{n-k-1}-E'_{n-k-1}\right),
\end{equation}
from which follow
\begin{eqnarray}
z_n&=&\rho^nz_0+\sum_{k=0}^{n-1}\rho^kE_{n-k-1}\nonumber\\
&+&{{1}\over{2}}\ \rho^n\left((1+2\alpha)^n-1\right)\xi_0+\alpha\sum_{k=1}^n\rho^k\sum_{l=0}^{n-k-1}\rho^l(1+2\alpha)^l\left(E_{n-k-l-1}-E'_{n-k-l-1}\right),\nonumber\\
z'_n&=&\rho^nz'_0+\sum_{k=0}^{n-1}\rho^kE'_{n-k-1}\nonumber\\
&+&{{1}\over{2}}\ \rho^n\left((1+2\alpha)^n-1\right)\xi_0-\alpha\sum_{k=1}^n\rho^k\sum_{l=0}^{n-k-1}\rho^l(1+2\alpha)^l\left(E_{n-k-l-1}-E'_{n-k-l-1}\right).
\label{eqn:z_n,z'_n}
\end{eqnarray}
The coupling dependence of the first two terms in $z_n$ and $z'_n$ enters only through the function $E$. If we ignore the detail inside of the disk of the size $R_\epsilon$, we do not distinguish these terms from the points of the Julia set $J^{(0,\epsilon)}$ of uncoupled system. 

The last two terms both of $z_n$ and $z'_n$ exhibit explicit dependence on the coupling $g$. The feature of the dependence is quite interesting and we will study further detail. The term ${{1}\over{2}}\ \rho^n\left((1+2\alpha)^n-1\right)\xi_0$ depends on the initial position ${\mbox{\boldmath{$Z$}}}_0$ explicitly. It tends to shift the central position $\left(\matrix{\rho^nz_0\cr \rho^nz'_0\cr}\right)$ on the `Julia set' $J^{(0,0)}=\left(\matrix{J^{ILM}\cr J^{ILM}\cr}\right)$ by a small amount as long as the coupling, hence $\alpha$, is small. The last term is a finer correction whose dependence on the initial position comes only through the function $E$. Using the inequality $(\ref{eqn:|E(z)|<})$ the value of this term can be estimated as
\begin{eqnarray}
&&\left|\alpha\sum_{k=1}^n\rho^k\sum_{l=0}^{n-k-1}\rho^l(1+2\alpha)^l\left(E_{n-k-l-1}-E'_{n-k-l-1}\right)\right|\nonumber\\
&&\quad\qquad <\ {2|\alpha| R_{\epsilon}\over 1-|\rho(1+2\alpha)|}\left(|\rho|{1-|\rho|^n\over 1-|\rho|}-|\rho|^n{|1+2\alpha|^n-1\over |1+2\alpha|-1}\right)
\end{eqnarray}
The value of the right hand side is determined only by the parameters characterizing the map. Therefore the information about the initial values of the map is lost in this term. 

It is remarkable that we could get an `analytic expression' of an orbit after multiple inverse map $M_{(1)}$. Since $A$ contains a square root function, it is not trivial at all to get any compact expression of an orbit for the coupled system. What is meant by `analytic expression' is that all expressions, but contents of the function $E$, are given explicitly. The maximum value of $E$ is bounded by $R_\epsilon$, irrespective to the position of the map. We loose informations inside of the disk at every step of the map. The uncertainty, however, does not increase as steps of the map increase, since the difference from the uncoupled case is calculated independently at every step of the map. This fact enables us to predict the inverse orbit, as far as detail informations inside of the disk are not required.

\subsubsection{Maps of general form}

If $\mbox{\boldmath{$X$}}$ is the image of ${\mbox{\boldmath{$Z$}}}_0$ by $M_{(j_1,j_2,\dots,j_{n-s-1})}$ with $s$ between 0 and $n-1$, $\{M_{(1)}^sM_{(j)}\mbox{\boldmath{$X$}};\ j=1,2,3,4\}$ represent general form of the map $M_{(j_1,j_2,\dots,j_n)}{\mbox{\boldmath{$Z$}}}_0$. We have already discussed the extreme case, $s=n-1,\ j=1$. From previous experience it is not difficult to see general feature of the map. 

The image of $M_{(j)}\mbox{\boldmath{$X$}}$ by the map $M_{(1)}^s$ is given by $(\ref{eqn:z_n,z'_n})$ if $n$ is replaced by $s$ and ${\mbox{\boldmath{$Z$}}}_0$ is replaced by those from $(\ref{eqn:M_(1)})$ to $(\ref{eqn:M_(4)})$. 
\begin{eqnarray}
z_{n,j}&=&\rho^sz_{n-s,j}+\sum_{k=0}^{s-1}\rho^kE_{s-k-1}+\rho^s\left((1+2\alpha)^s-1\right)\xi_{n-s,j}\nonumber\\
&+&\alpha\sum_{k=1}^s\rho^k\sum_{l=0}^{s-k-1}\rho^l(1+2\alpha)^l\left(E_{s-k-l-1}-E'_{s-k-l-1}\right),\nonumber\\
z'_{n,j}&=&\rho^sz'_{n-s,j}+\sum_{k=0}^{s-1}\rho^kE'_{s-k-1}+\rho^s\left((1+2\alpha)^s-1\right)\xi_{n-s,j}\nonumber\\
&-&\alpha\sum_{k=1}^s\rho^k\sum_{l=0}^{s-k-1}\rho^l(1+2\alpha)^l\left(E_{s-k-l-1}-E'_{s-k-l-1}\right),
\label{eqn:z_s,z'_s}
\end{eqnarray}
where $\left(\matrix{z_{n-s,j}\cr z'_{n-s,j}\cr}\right)=M_{(j)}\mbox{\boldmath{$X$}}$. We notice the main difference of this expression from the one of $(\ref{eqn:z_n,z'_n})$ is that the first terms $\rho^nz_0$ and $\rho^nz'_0$ in $(\ref{eqn:z_n,z'_n})$ are points of the Julia set $J^{ILM}$ of uncoupled system, whereas $\rho^sz_{n-s,j}$ and $\rho^sz'_{n-s,j}$ are not in general. If the last operation of $M_{(j)}$ was $M_{(4)}$, they must stay in the neighbourhood of $-\rho^{n-s}\lambda$, a point in $J^{ILM}$. Hence $\left(\matrix{z_{n,4}\cr z'_{n,4}\cr}\right)$ also remain in the neibourhood of $J^{(0,\epsilon)}$. Otherwise we do not have full information. 

If there exist points, whose locations relative to those of uncoupled case are not known, the Julia set $J^{(g,\epsilon)}$ of the coupled system tends to spread compared with $J^{(0,\epsilon)}$. Detail inspection of the expression $(\ref{eqn:z_s,z'_s})$, however, will show that such expansion of the Julia set will not continue without limit, as long as the parameter $\rho$ is chosen less than 1 and the coupling strength $g$ is sufficiently small. In fact terms including $z_{n-s}$ or $\xi_{n-s}$, hence dependent on the locations, are multiplied by factors of the form $\rho^s$ and/or $(1+2\alpha)^s-1$, which tend to push the points to the neighbourhood of the origin. The behaviour of the map will change its nature significantly depending if the value of $|\rho(1+2\alpha)|$ is greater than 1 or less. The critical value is given by
\begin{equation}
g_c={1\over 2}\ {1-|\lambda'|\over 1+|\lambda'|}.
\end{equation}

The change of the Julia set with the coupling parameter $g$ is represented below. The simulations are executed at $\lambda=4$ and $\lambda'=-0.3$. In this case $|\Lambda_\infty|\le 1$ and $|\Lambda_0| > 1$ hold for small $g$, hence the Julia set exists in a finite region including the origin. This condition, however, is destroyed at some critical value of $g$ since the condition of the convergence depends on both $|\Lambda_\infty|$ and $|\Lambda_0|$ and the Julia set may spread to infinity. In our choice of the parameters we observe
\begin{enumerate}
\item $g={1-\sqrt{|\lambda'|}\over 2}\approx 0.226$

One of $|\Lambda_\infty(r=-1)|$'s becomes 1, hence the Julia set starts to includes the infinity.
\item $g=g_c\approx 0.269$

$|\rho(1+2\alpha)|$ becomes 1, hence the distance from the uncoupled case fails to be estimated.

\item $g={1-|\lambda'|\over 2}=0.35$

Another $|\Lambda_\infty(r=-1)|$ becomes 1, hence the Julia set tends to expand to infinity.

\item $g={1\over 2}\left(1-{1\over |\lambda|}\right)=0.375$

$|\Lambda_0|$ becomes 1, hence the origin ceases to be a repeller and the Julia set starts to cover inside of the basin of $z_p$.
\end{enumerate}

Here we like to visualize the coupling dependence of the Julia set by using the computer simulation of the inverse map when other parameters are fixed. Like the previous pictures only projections of the Julia set are visible. In the present case we will use the projections to the $z$ plane, so that we can compare the influence of the coupling with those of single map. The size of the windows is $8\times 8$ and the center of the windows is at $(Re(z),\ Im(z))=(-2,0)$ of the complex plane. 

\vfill\eject
\begin{minipage}{16cm}
\vglue 0.5cm
\noindent
{\bf Fig.2 :Julia set projected on $z$-plane}
\hfill\break
\vglue 0.2cm
\noindent
\begin{minipage}{7cm}
Fig.2a :$g=0.00$\\
\epsfxsize=7cm \epsfbox{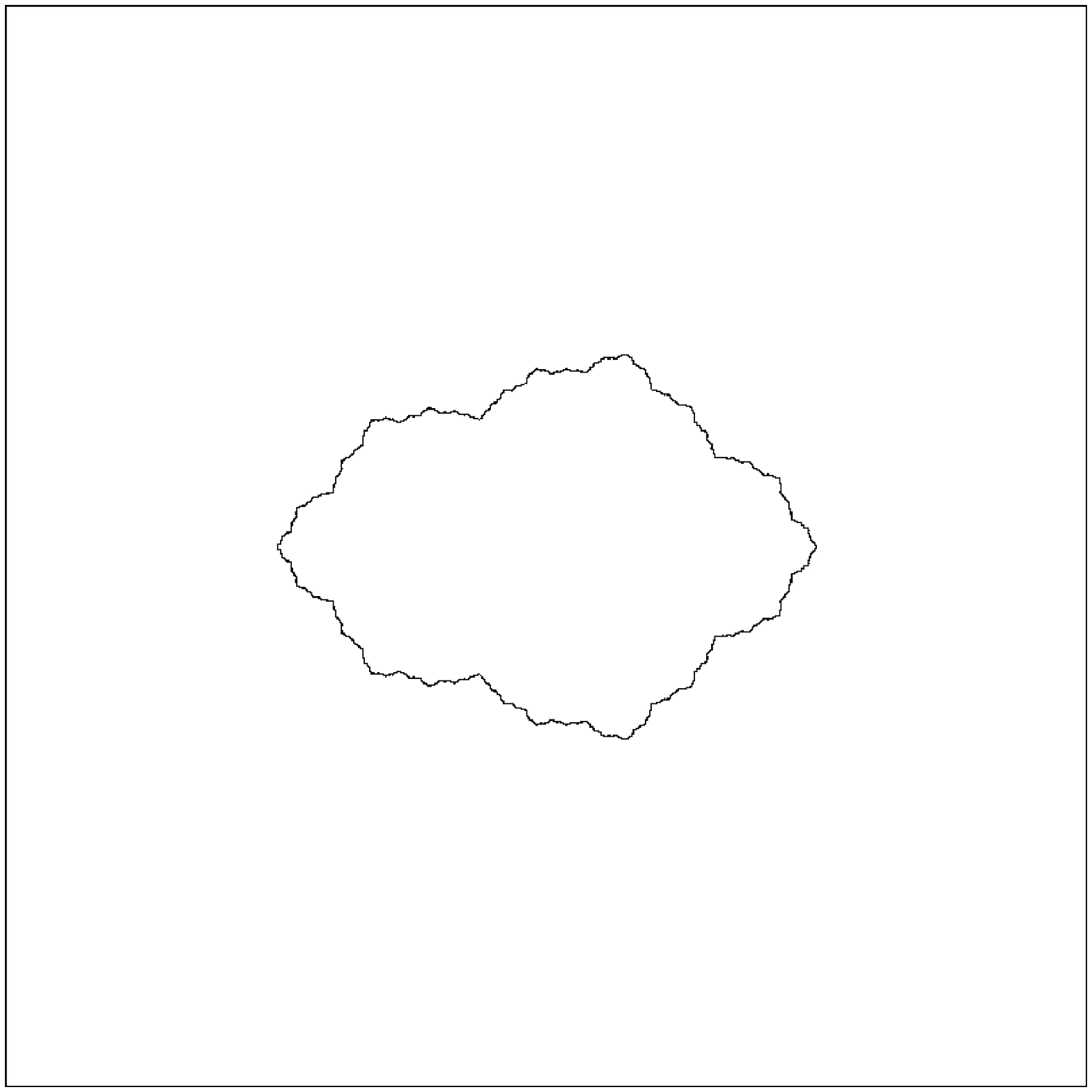}
\end{minipage}
\hfill
\begin{minipage}{7cm}
Fig.2b :$g=0.20$\\
\epsfxsize=7cm \epsfbox{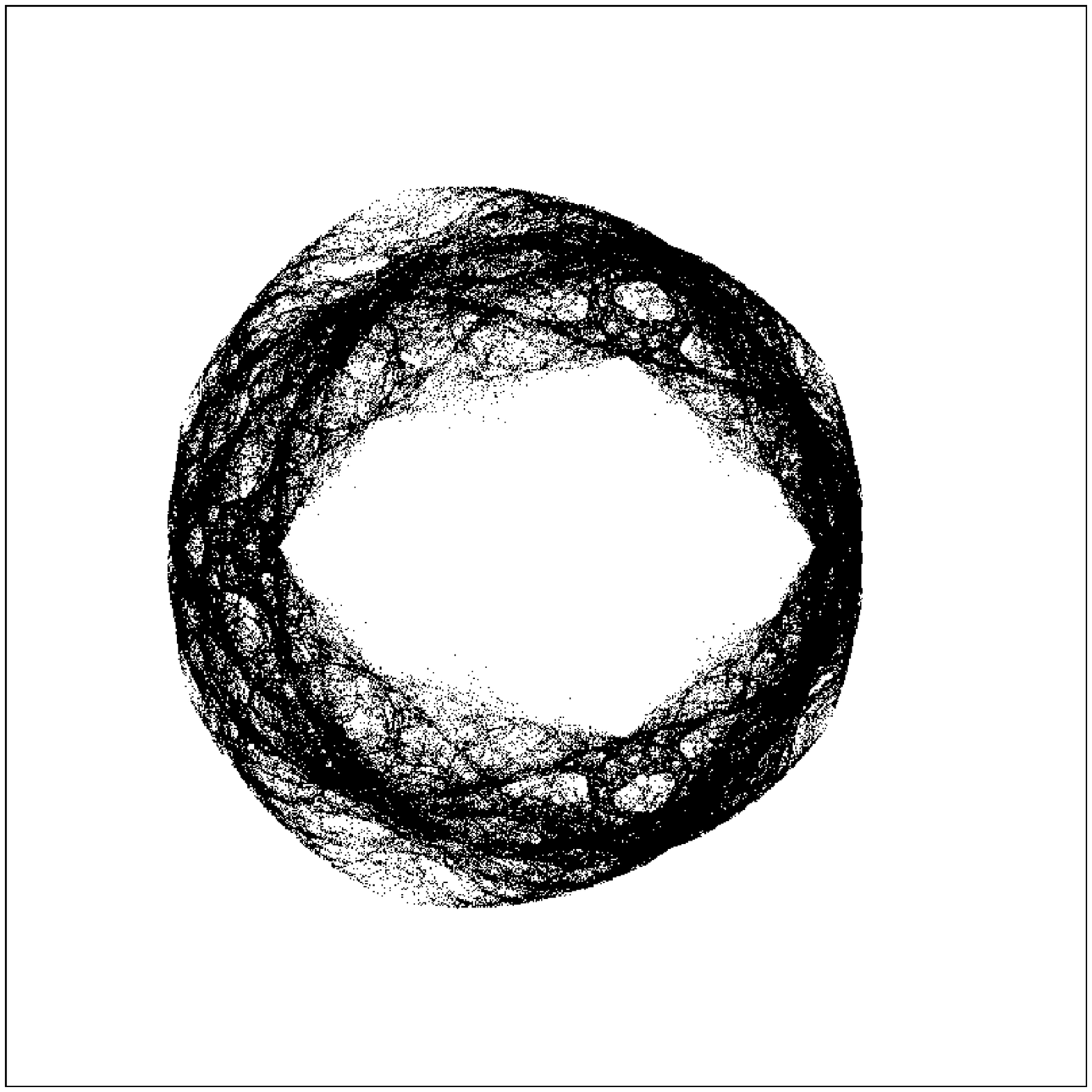}
\end{minipage}
\hfill
\\
\noindent
\hfill
\begin{minipage}{7cm}
Fig.2c :$g=0.25$\\
\epsfxsize=7cm \epsfbox{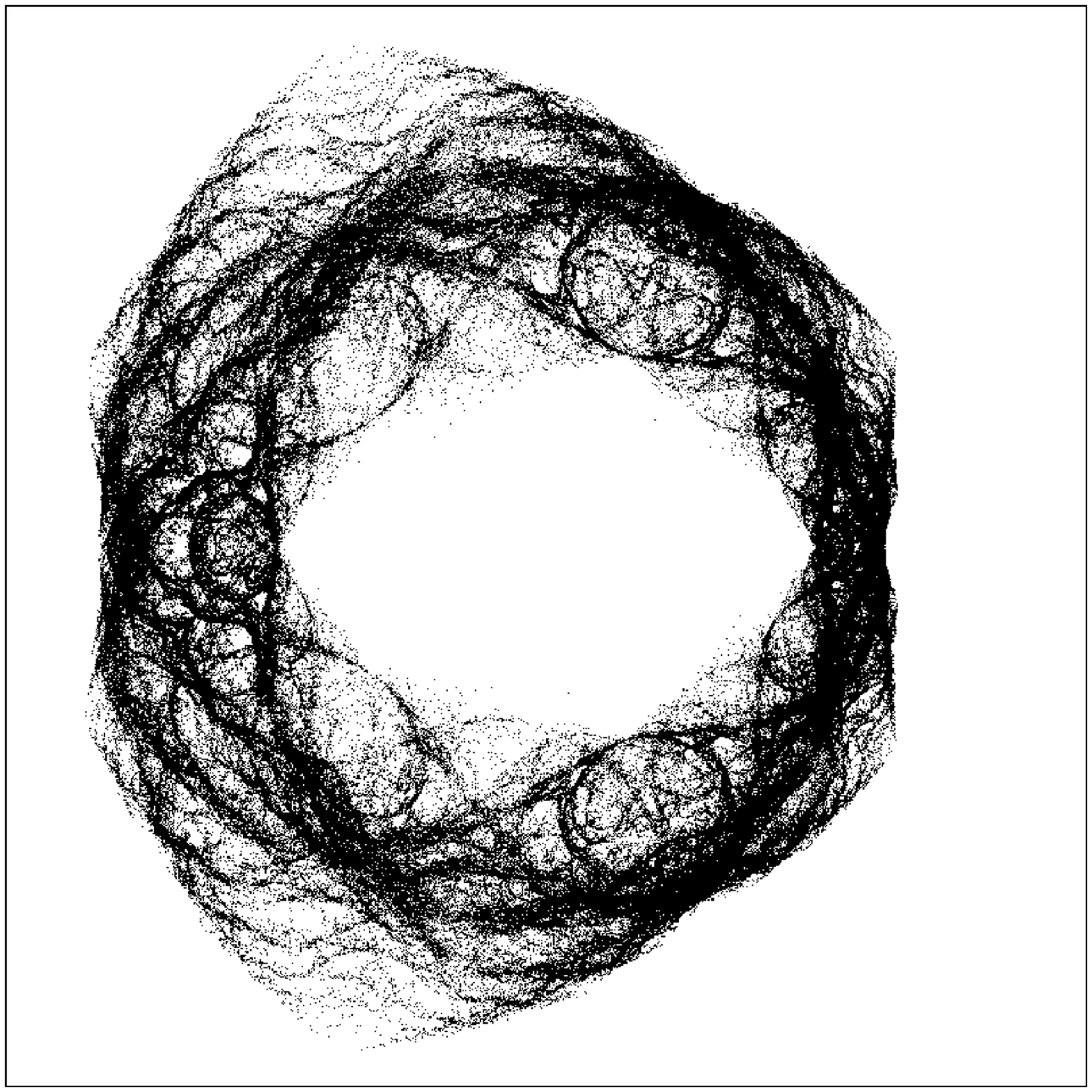}
\end{minipage}
\hfill
\begin{minipage}{7cm}
Fig.2d :$g=0.30$\\
\epsfxsize=7cm \epsfbox{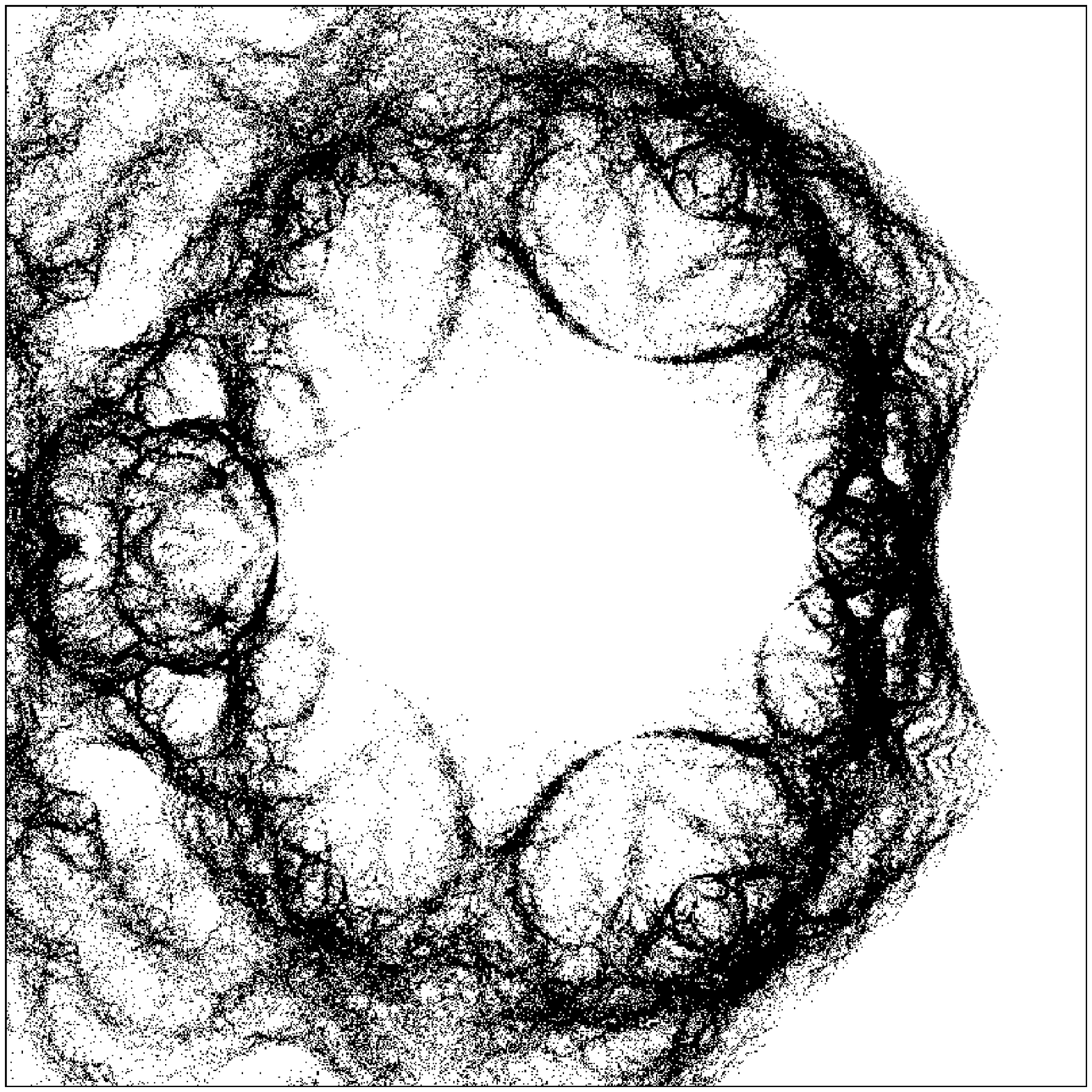}
\end{minipage}
\hfill
\\
\noindent
\hfill
\begin{minipage}{7cm}
Fig.2e :$g=0.35$\\
\epsfxsize=7cm \epsfbox{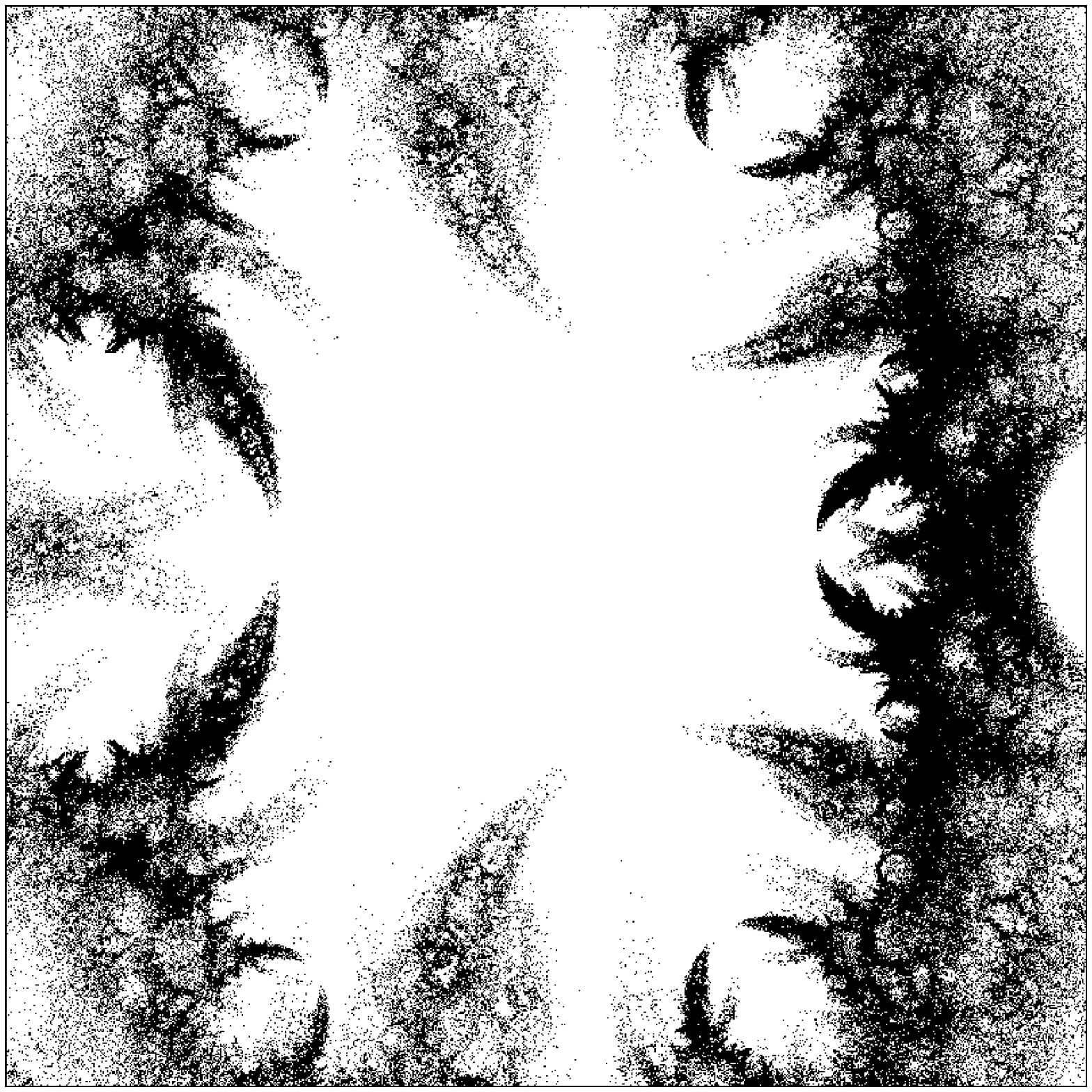}
\end{minipage}
\hfill
\begin{minipage}{7cm}
Fig.2f :$g=0.40$\\
\epsfxsize=7cm \epsfbox{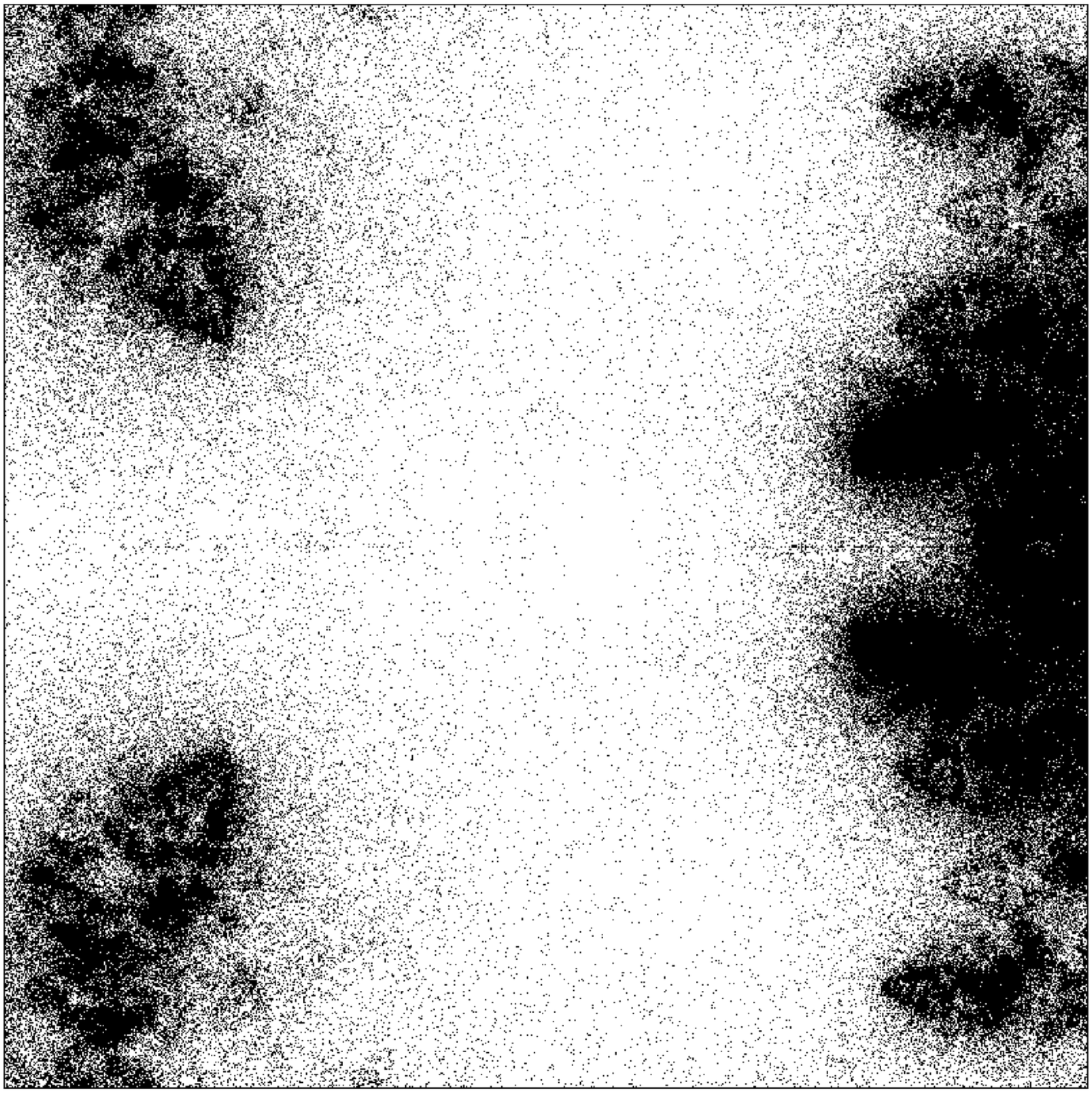}
\end{minipage}
\hfill
\end{minipage}

\vfill\eject
\section{Discussions}

We have investigated a coupled complex dynamical system which is generated by degree two rational map. The subject we considered in this study was the Julia set. It characterizes dynamical feature of the system. It is an invariant object in the sense that it does not depend on initial values. It, however, changes very sensitively if the parameters of the map are changed. In particular we were interested in how the Julia set is influenced when two independent maps are coupled. We have chosen the model of the coupling carefully, so that we can study the effect of the coupling, not only numerically, but analytically to some extent. The introduction of the coupling tends to spread the Julia set when it is projected to the complex plane of the single map. 
This naturally happens because the projection procedure linearly superposes the effects of the other unit of the map. As the coupling strength increases beyond the critical value $g_c$, the Julia set extends to infinity. This was observed also in the numerical calculations.

Another problem which we were interested in is concerned with the transition of the system from chaotic phase to integrable phase. It is a hard task in general to signify such a transition, because it requires to analyze some singular functions which appear in the inverse map. When the system is complex the inverse functions are seldom to be studied analytically. Therefore it is remarkable that we could analyze the transition of the Julia set of our coupled system and could prove the uniform transition to the integrable phase.

Before closing this paper we will discuss briefly an attempt of measuring complexity of the Julia set. For this purpose we study fractal dimensions of the Julia set. Since the Julia set is an invariant of the map, the fractal dimensions are also invariant once the parameters of the map are fixed. We are interested in how the fractal dimensions change as the coupling between two GLM's varies or as the integrability parameter $\epsilon$ approaches to its critical value.

There are several kinds of definition called fractal dimension. We will use the box dimension and the information dimension in the following, which are relatively easy to calculate from the data of computer simulation.

The box dimension is defined by
$$
d_B = \lim_{\delta \rightarrow 0}{{ \log( N(\delta) )}\over{\log(1/\delta)}}  
$$
where $N(\delta)$ denotes a number of points which constitute the figure under consideration and $\delta$ represents the length which decides minimum scale of the measure.  

The information dimension is defined by
$$
d_I = \lim_{\delta \rightarrow 0}{{ \log( S(\delta) )}\over{\log(1/\delta)}}  
$$
in which $S$ is the entropy obtained from the probability $P_j$ that the mapping across the $j$th-cell box of length $\delta$,
$$
S(\delta) \equiv - \sum_{j = 0}^{\infty} P_j\log(P_j).
$$

In the calculation of the fractal dimensions there arise some problems due to the use of computer. Namely, since we can not take the limit $\delta \rightarrow 0$ in the computer, there could remain some ambiguities on the values of the calculation. We performed the calculation at different values of $\delta$ and found that it is enough if the value is sufficiently small. To be specific we will fix the minimum scale $\delta$ to be ${{1}\over{800}}$ of the window length in each computer simulation. In this ratio, we calculate 800,000 points for each plot of fractal dimension with a set of parameters.
 This number of points is determined such that the dimension of the random mapping is sufficiently close to 2.

Another problem which arises in our calculation is due to the fact that the Julia set is an object in two dimensional complex, hence four dimensional real space. Very huge number of points are needed to calculate the fractal dimensions. In order to reduce the number of calculations we performed the calculation only of projections of the Julia set to two dimensional real space. In other words we abandon half of the informations about location of each point in the four dimensional space. The values obtained in this way do not mean those in the full dimensions. Nevertheless the feature of their variation should represent characterization of structure possessed by the Julia sets with different values of parameters. 

The results of our computer calculation of the fractal dimension of the Julia set are shown in Figs. 3. The first two figures a, b are for the fixed values of $\lambda=4$ and $g=0.3$, whereas the integrability parameter $\epsilon$ is changed from $-1$ to $0$. They exhibit the gradual decrease of the dimension as the integrability parameter $\epsilon$ approaches to the integrable limit, hence agree with the analytical result. The same quantities were also calculated against the coupling parameter $g$. 

Figs. 3 c, d present the results of the calculation for various values of the coupling $g$ between 0 and 0.45, while $\lambda$ and $\lambda'$ are fixed at 4 and $-0.3$. We observe that both box and information dimensions increase as the coupling increases when $g$ is small. Beyond the critical value $g_c$ ( $= {{7}\over{26}}$ in the case of the calculation), we observe no significant change of the values of the fractal dimension. This means that the expansion of the Julia set to the whole space takes place gradually as the coupling of the system becomes strong beyond the critical value. However such an expansion, in turn, causes disappearence of the points from our observation.

%
\begin{center}
\begin{minipage}{11cm}
\epsfxsize=10cm \epsfbox{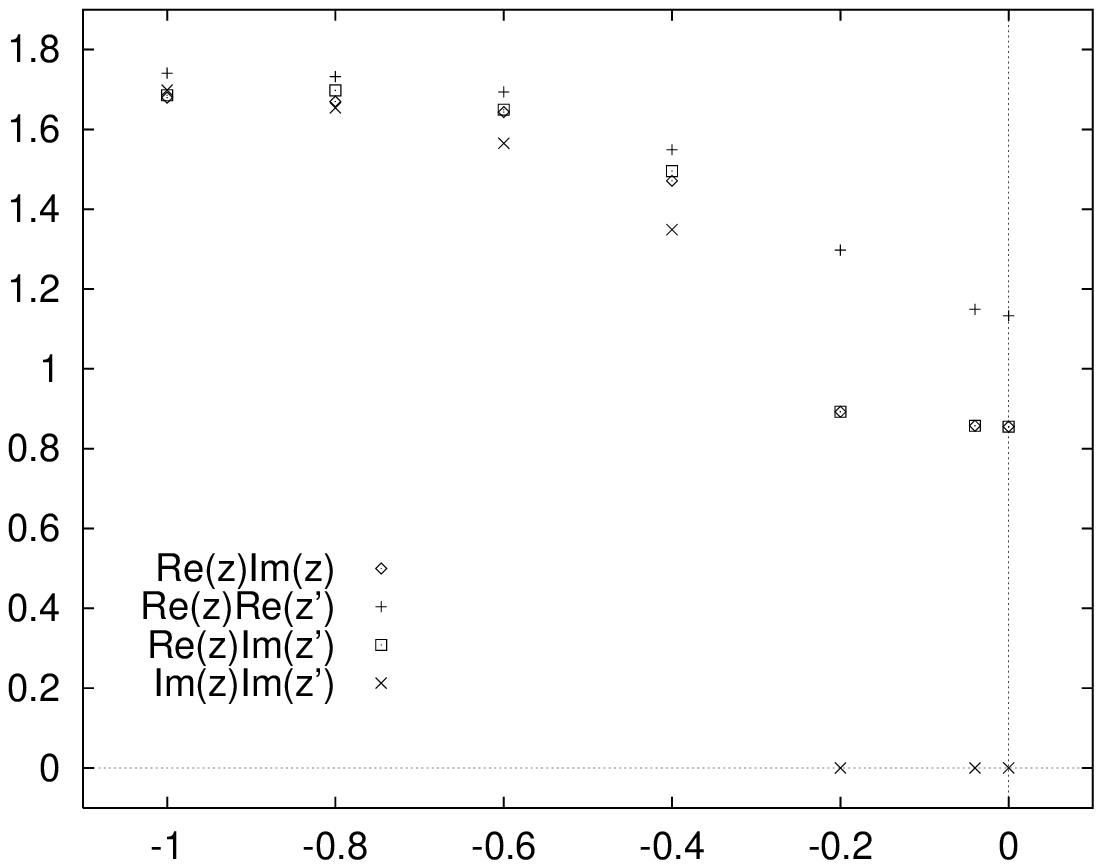}
{\bf Fig.3a Box dimension vs the parameter $\epsilon$:}\hfill\break
\end{minipage}
\end{center}

%
\begin{center}
\begin{minipage}{11cm}
\epsfxsize=10cm \epsfbox{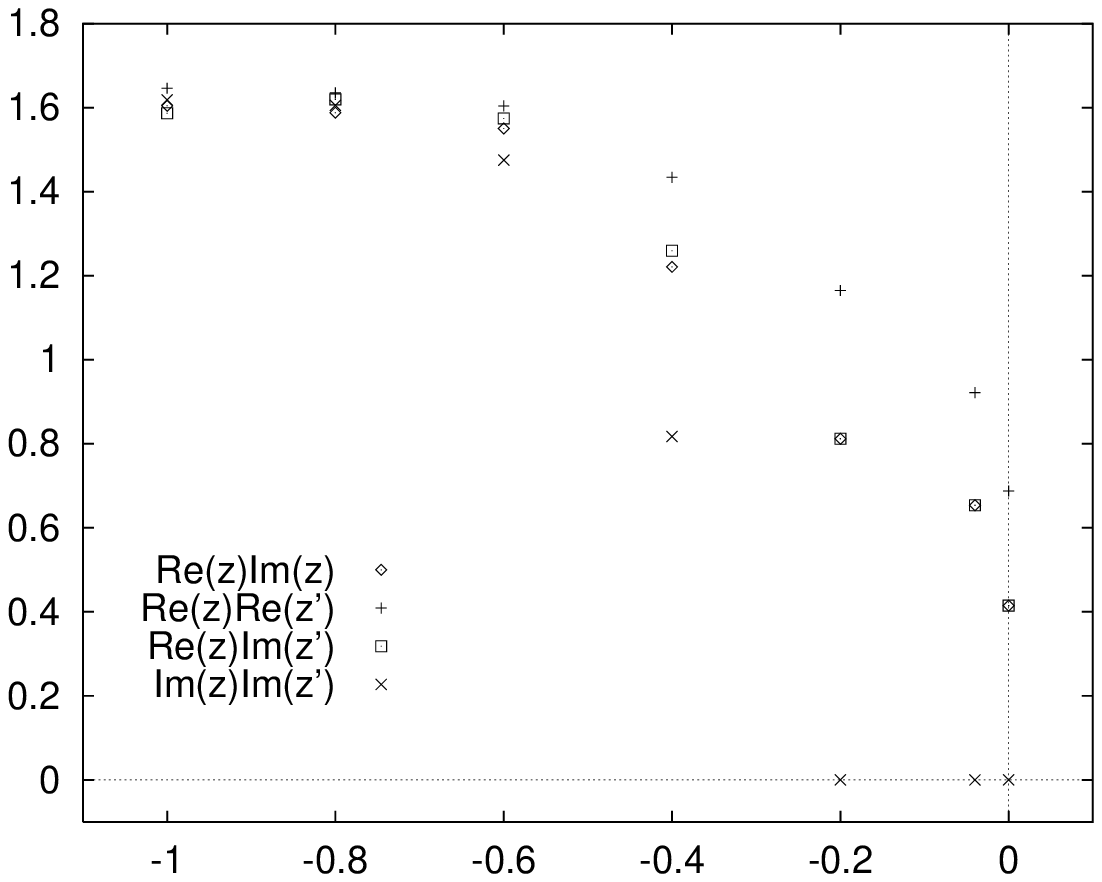}
{\bf Fig.3b Information dimension vs the parameter $\epsilon$:}\hfill\break
\end{minipage}
\end{center}
%
\begin{center}
\begin{minipage}{11cm}
\epsfxsize=10cm \epsfbox{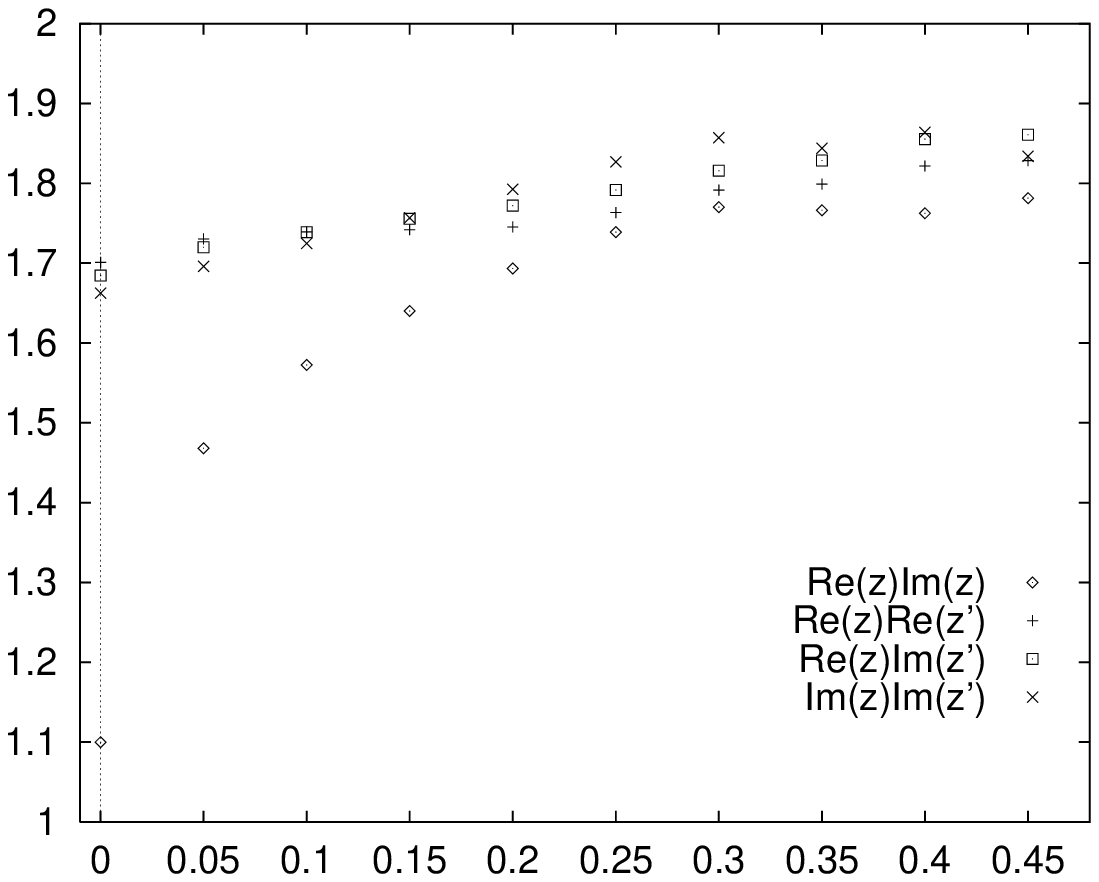}
{\bf Fig.3c Box dimension vs the parameter $g$:}\hfill\break
\end{minipage}
\end{center}

%
\begin{center}
\begin{minipage}{11cm}
\epsfxsize=10cm \epsfbox{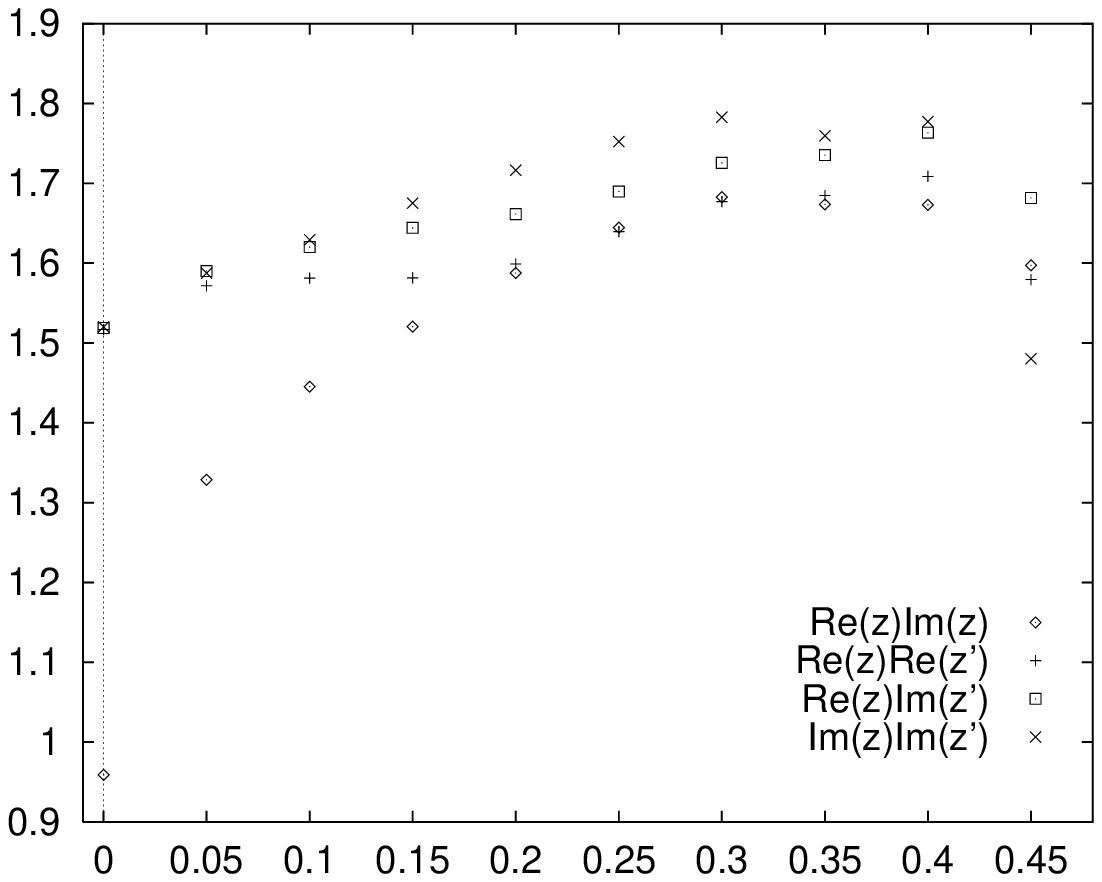}
{\bf Fig.3d Information dimension vs the parameter $g$:}\hfill\break
\end{minipage}
\end{center}




\newpage
\noindent {\Large\bf Appendix}

The Jacobi matrix of the coupled system is defined by
\begin{equation}
\left(\matrix{{\partial\varphi(z,z')\over\partial z}&{\partial\varphi(z,z')\over\partial z'}\cr {\partial\varphi(z',z)\over\partial z}&{\partial\varphi(z',z)\over\partial z'}}\right)
\ =\ 
\left( 
\begin{array}{cc}
 (1-g) F'( z ),& g F'( z' )\\
  g F'( z ),& (1-g) F'( z' )
\end{array}
\right)
\end{equation}
where 
\begin{equation}
F'( z ) = {\frac{d F( z )}{dz}} = {\frac{ \lambda' z^2 + 2 z + \lambda }{ ( 1 + \lambda' z )^2 }}e^{i\theta}.
\label{eqn:X(z)}
\end{equation}
The eigenvalues of this matrix are
$$
\Lambda = \frac{1}{2} (1-g) ( F'( z ) + F'( z' ) ) \pm \frac{1}{2} \sqrt{ (1-g)^2 ( F'( z ) - F'( z' ) )^2 + 4 g^2 F'( z ) F'( z' )},
$$

The behaviour of the dynamical map at infinity is determined by the Jacobi matrix
\begin{equation}
\left(\matrix{{\partial\over\partial\xi}{1\over \varphi\left({1\over\xi},{1\over\xi'}\right)}&
{\partial\over\partial\xi}{1\over \varphi\left({1\over\xi'},{1\over\xi}\right)}\cr
{\partial\over\partial\xi'}{1\over \varphi\left({1\over\xi},{1\over\xi'}\right)}&
{\partial\over\partial\xi'}{1\over \varphi\left({1\over\xi'},{1\over\xi}\right)}}\right)
\end{equation}
which should be calculated at $\xi={1\over z}=0,\ \xi'={1\over z'}=0$.

In order to get idea about stability of the map near the fixed points, we calculated regions of parameters which correspond to repulsive, attractive and saddle points. Under the restrictions of the the parameters by $\lambda'=\lambda$ and $\theta=0$, we can draw the picture on the $(g,\ \lambda)$ plane. Figs. a,b,and c show the parameter plane associated with the fixed points at $\mbox{\boldmath $Z$}_0$, $\mbox{\boldmath $Z$}_p$, and $\mbox{\boldmath $Z$}_\pm$, respectively\footnote{The case of $\mbox{\boldmath $Z$}_{\infty}(r=1)$ is the same with one of $\mbox{\boldmath $Z$}_0$.}. Here the most dark gray region represents the parameters corresponding to repulsive fixed points, the gray to attractive and the lightest one to saddle points.

\begin{minipage}{5cm}
\epsfxsize=5cm \epsfbox{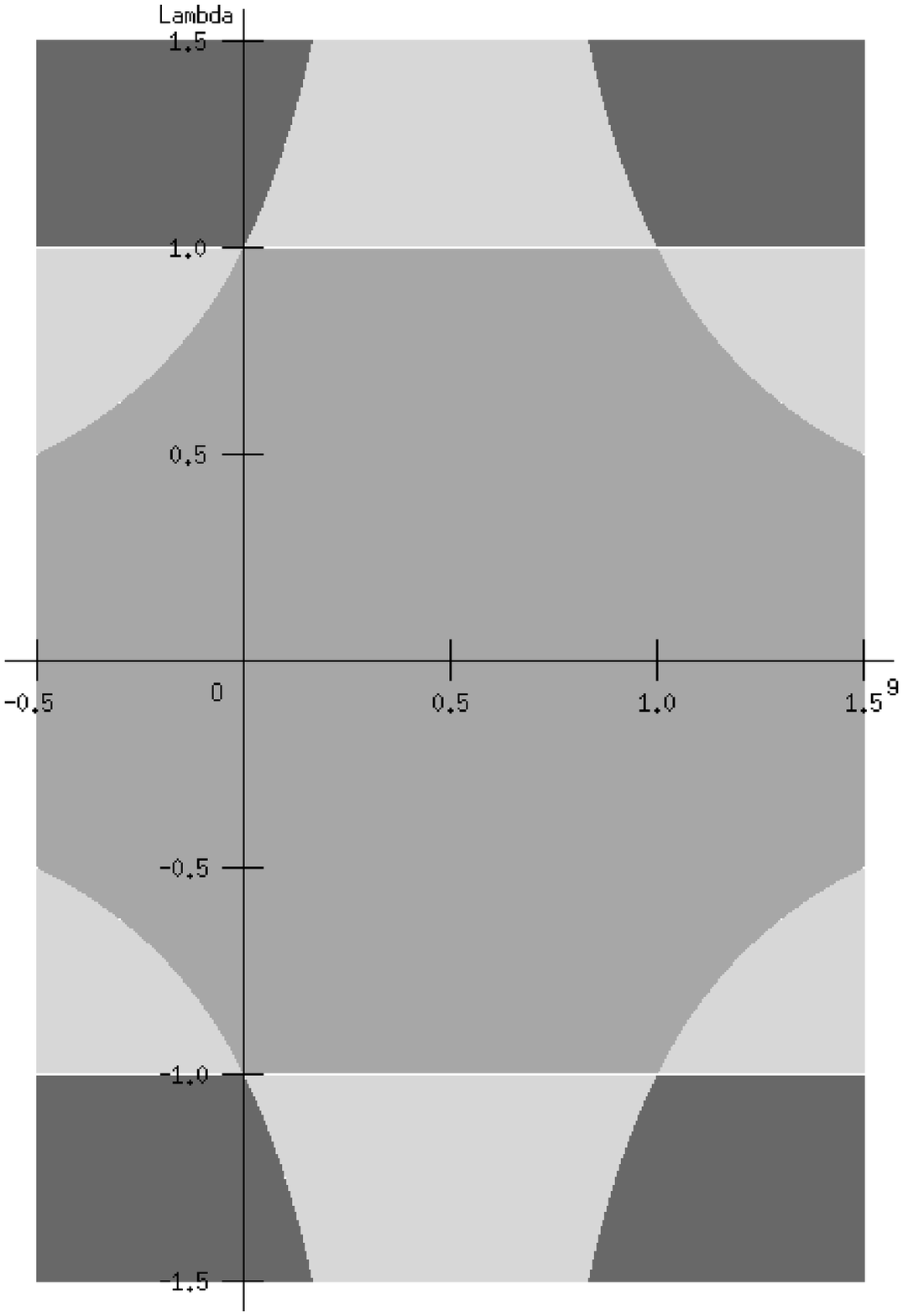}
Fig.a
\end{minipage}
\begin{minipage}{5cm}
\epsfxsize=5cm \epsfbox{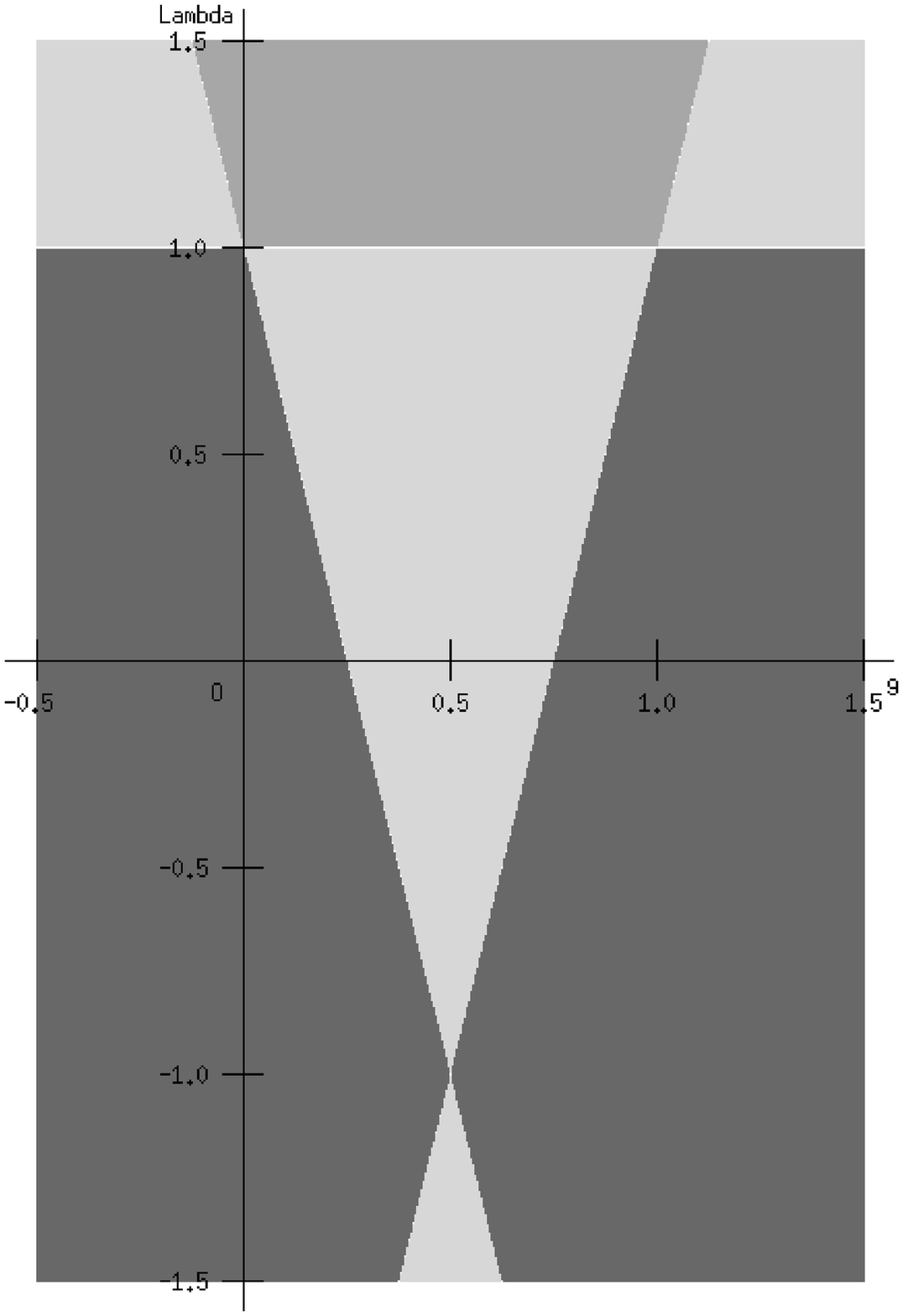}
Fig.b
\end{minipage}
\begin{minipage}{5cm}                   
\epsfxsize=5cm \epsfbox{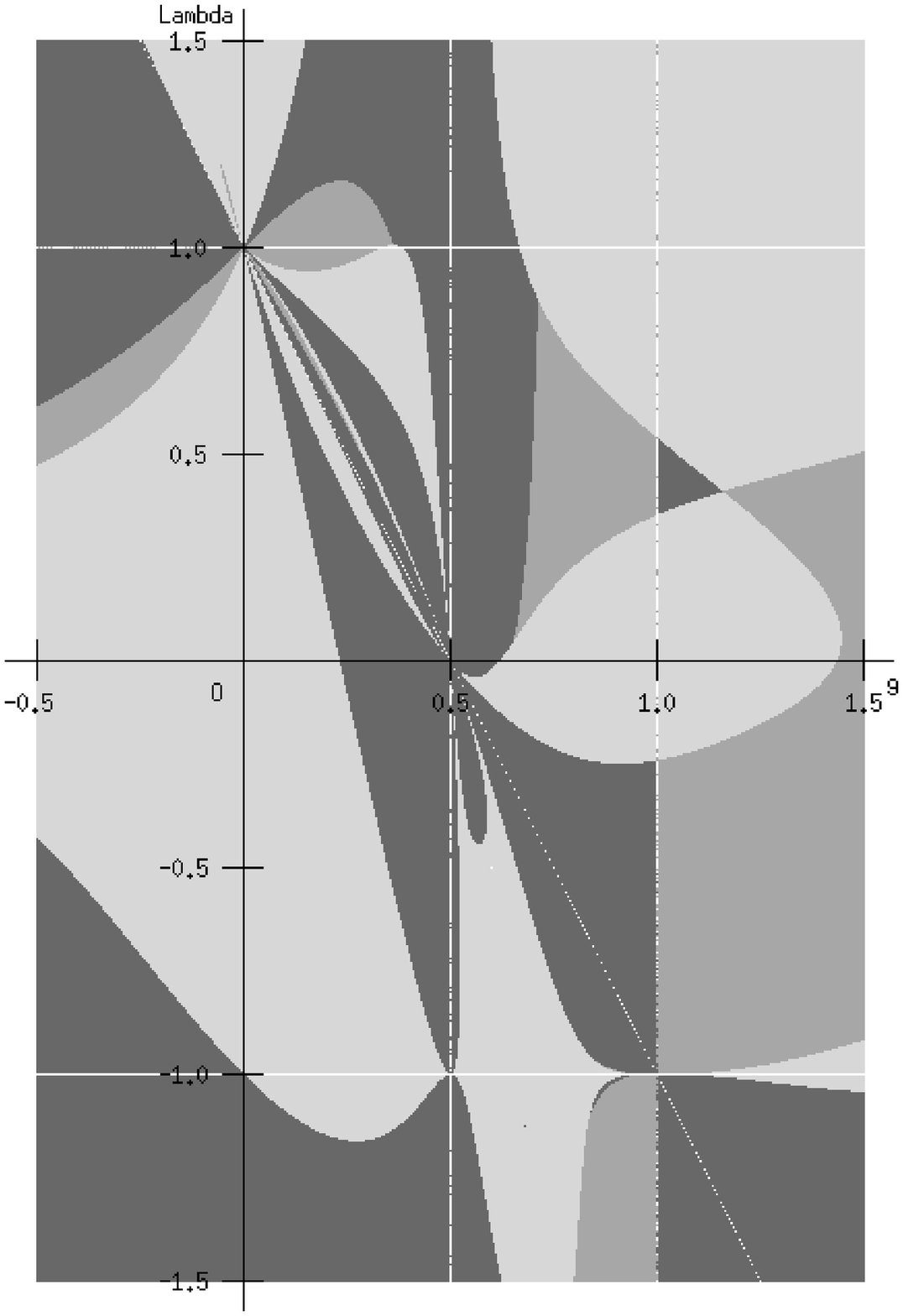}
Fig.c
\end{minipage}

\end{document}